\documentclass[twocolumn,english,aps,prb,floatfix,preprintnumbers,showpacs,amsfonts,amssymb,superscriptaddress]{revtex4}
\usepackage{ae,aecompl}
\usepackage[T1]{fontenc}
\usepackage[latin1]{inputenc}
\usepackage{color}
\usepackage{float}
\usepackage{booktabs}
\usepackage{amsmath}
\usepackage{amssymb}
\usepackage{graphicx}

\makeatletter

\providecommand{\tabularnewline}{\\}

\@ifundefined{textcolor}{}
{%
 \definecolor{BLACK}{gray}{0}
 \definecolor{WHITE}{gray}{1}
 \definecolor{RED}{rgb}{1,0,0}
 \definecolor{GREEN}{rgb}{0,1,0}
 \definecolor{BLUE}{rgb}{0,0,1}
 \definecolor{CYAN}{cmyk}{1,0,0,0}
 \definecolor{MAGENTA}{cmyk}{0,1,0,0}
 \definecolor{YELLOW}{cmyk}{0,0,1,0}
}

\usepackage{amscd}
\usepackage{bm}
\usepackage{graphics,psfrag}
\usepackage{graphicx,psfrag}
\usepackage{lipsum}

\makeatother

\usepackage{babel}
\begin{document}
\title{Coexistence of spontaneous dimerization and magnetic order in a transverse-field
Ising ladder with four-spin interactions }
\author{J.~C.~Xavier }
\affiliation{Universidade Federal de Uberlândia, Instituto de F\'{\i}sica, C. P
593, 38400-902 Uberlândia, MG, Brazil }
\author{R.~G. Pereira}
\affiliation{International Institute of Physics and Departamento de Física Teórica
e Experimental,Universidade Federal do Rio Grande do Norte, 59072-970
Natal-RN, Brazil }
\author{M.~E.~S. Nunes }
\affiliation{Universidade Federal de Ouro Preto, Departamento de Física, ICEB,
Campus Universitário Morro do Cruzeiro, 35400-000 Ouro Preto, MG,
Brazil}
\author{J.~A. Plascak}
\affiliation{Universidade Federal da Paraíba, Departamento de Física, CCEN, Campus
I, s/n, Cidade Universitária, 58051-090 João Pessoa, PB, Brazil}
\affiliation{Departamento de Física, Instituto de Ciências Exatas, Universidade
Federal de Minas Gerais, C.P. 702, 30123-970 Belo Horizonte, MG -
Brazil }
\affiliation{University of Georgia, Department of Physics and Astronomy, 30602
Athens, GA, USA}
\date{\today{}}
\begin{abstract}
The spin-1/2 transverse field two-leg Ising ladder with nearest-neighbor
exchange and plaquette four-spin interaction $J_{4}$ is studied analytically
and numerically with the density matrix renormalization group approach.
The quantum phase diagram in the transverse field $B$ versus $J_{4}$
plane has been obtained. There are three different phases: a paramagnetic
(PM) phase for high values of the transverse field and, for low values
of $B$, a ferromagnetic (FM) ordered phase for small $J_{4}$ and
a dimerized-rung (DR) phase for large negative values of $J_{4}$.
All phases are separated by quantum phase transition lines meeting
at a multicritical point. The critical lines have been obtained by
exploring the entanglement entropy. The results show that along the
critical lines the central charge is $c=1/2$, while at the multicritical
point one has $c=1$. The scaling dimension of the energy operator
is $X_{\epsilon}=1$, in agreement with the universality class of
the critical behavior of the quantum Ising chain. An effective field
theory for the multicritical point is also discussed. The FM and the
DR order parameters have also been computed and we found a region
where the FM and the DR phases coexist. 
\end{abstract}
\maketitle

\section{INTRODUCTION}

The topic of quantum spin systems is by far one of the most fascinating
and interesting branches of experimental and theoretical physics.
Although in the beginning the study of quantum spin systems was specifically
related to magnetism and the corresponding magnetic properties \citep{park,rat},
it has by now migrated to several areas such as high-temperature superconductors
\citep{RevModPhys.78.17}, ferromagnetic nanowires \citep{cho}, and
spintronics \citep{RevModPhys.76.323}, among others. Quantum phase
transitions have also been recently revisited and a close analogy
to the simple fluid phase diagram has been experimentally detected
in the geometrically frustrated quantum antiferromagnet SrCu$_{2}$(BO$_{3}$)$_{2}$
\citep{jim}. 

Low-dimensional quantum spin models with competing interactions represent
a fertile ground for unconventional magnetic behavior. Examples in
one dimension include the transverse Ising model with next-nearest-neighbor
interactions \citep{cola} and with linear interaction among four
spins \citep{bia,ozi1,ozi2}, ladder models describing cuprates \citep{hoh,nas},
and spin models with four-spin interactions used to explain ferroelectrics
\citep{jab,chu,teng}. Zero-dimensional models are also important
when applied to magnetic molecules and nanomagnetism \citep{kowa,jordana,karol},
multiferroics \citep{thomas,val}, and ultracold atoms trapped in
optical lattices \citep{kar,liu}.

While nearest-neighbor exchange is often the dominant interaction
in Mott insulating materials, longer-range and multi-spin interactions
may play an important role, particularly in the vicinity of a metal-insulator
transition \citep{macdonald,motrunich}. In fact, the ring exchange
interaction has been invoked to reproduce the dispersion relation
observed in inelastic neutron scattering experiments on cuprates such
as La$_{2}$CuO$_{4}$ and La$_{6}$Ca$_{8}$Cu$_{24}$O$_{41}$ \citep{brehmer,mat,col,lar}.
More recently, four-spin interactions have been argued to stabilize
a chiral spin liquid on the triangular lattice \citep{Cookmeyer}.

Motivated by experiments on materials with a ladder-like structure
\citep{chaboussant,watson,hong,schmidiger,blosser}, spin ladder models
have been extensively studied and shown to display rich phase diagrams.
For instance, the frustrated antiferromagnetic spin ladder in a magnetic
field exhibits a magnetization plateau at half saturation which is
attributed to a gapped state that spontaneously breaks the lattice
translational symmetry \citep{cabraetal0,honecker2000,fouet}. A columnar
dimer phase was also proposed at zero magnetic field \citep{Starykh2004,Hikihara},
but the numerical evidence for this phase is still ambiguous \citep{Barcza}.
Moreover, four-spin interactions in spin ladders can give rise to
other unusual types of order, such as scalar chirality and intraleg
staggered dimerization \citep{sakai,lauchli,calzado,gritsev,liu2008,caponi-nleg}. 

The purpose of the present work is to study the effects of four-spin
interactions on the transverse-field Ising model defined on a two-leg
ladder. Based on analytical considerations and density matrix renormalization
group (DMRG) methods, we map out the phase diagram of the model and
characterize the nature of the transitions. In the regime of dominant
four-spin interaction, we find a non-magnetic phase which breaks translational
invariance by means of rung dimerization. Remarkably, for intermediate
values of the four-spin interaction this dimer order coexists with
the conventional ferromagnetic (FM) order of the quantum Ising model
at weak fields. Analyzing the entanglement entropy, we identify two
critical lines in the Ising universality class, with central charge
$c=1/2$. These lines cross at a multicritical point with central
charge $c=1$, below which the coexistence phase appears. 

The paper is organized as follows. In Sec. \ref{sec:MODEL} we introduce
the model and discuss the phases for some simple limits of the parameters.
The determination of the critical behavior, using the entanglement
entropy and the DMRG procedure, is described in Sec. \ref{sec:METHODOLOGY}.
Our main results are presented in Sec. \ref{sec:RESULTS}. Finally,
some concluding remarks are left for Sec. \ref{sec:CONCLUSIONS}.

\section{MODEL and phases\label{sec:MODEL}}

Consider the following Hamiltonian for a two-leg quantum Ising ladder
with periodic boundary conditions (PBC):
\begin{align}
{\cal H}= & -\sum_{n=1,2}\sum_{i=1}^{L}\left(\sigma_{n,i}^{z}\sigma_{n,i+1}^{z}+B\sigma_{n,i}^{x}\right)\nonumber \\
 & -\sum_{i=1}^{L}\left(J_{\perp}^{x}\sigma_{1,i}^{x}\sigma_{2,i}^{x}+J_{\perp}^{z}\sigma_{1,i}^{z}\sigma_{2,i}^{z}\right)\nonumber \\
 & -J_{4}\sum_{i=1}^{L}\sigma_{1,i}^{z}\sigma_{1,i+1}^{z}\sigma_{2,i}^{z}\sigma_{2,i+1}^{z}.\label{Model}
\end{align}
Here $\sigma_{n,i}^{\eta}$ ($\eta=x,z,y)$ are the Pauli spin matrices
at the \emph{i-}th site of leg \emph{$n=1,2$} and $L$ is the length
of the ladder, which has a total of $N=2L$ sites. The first term
with coupling constant set to unity corresponds to the Ising interaction
along the legs. In addition to a transverse magnetic field $B$, the
Hamiltonian includes rung couplings $J_{\perp}^{x}$ and $J_{\perp}^{z}$
in the $x$ and $z$ spin directions, respectively, as well as a four-spin
interaction $J_{4}$ on square plaquettes. For $J_{\perp}^{x}=J_{\perp}^{z}=J_{4}=0$,
the model reduces to two decoupled transverse-field Ising chains.
The well known quantum Ashkin-Teller model corresponds to $J_{\perp}^{x}=BJ_{4}$
and $J_{\perp}^{z}=0$ \citep{kohmoto,quantumAsh-Teller}. The latter
has a $\mathbb{Z}_{2}\times\mathbb{Z}_{2}$ symmetry generated by
$\mathcal{R}_{n}=\prod_{i}\sigma_{n,i}^{x}$ with $n=1,2$, equivalent
to global $\pi$ rotations around the $x$ axis for each leg independently.
In the complete model, the $J_{\perp}^{z}$ interaction breaks this
symmetry down to a single $\mathbb{Z}_{2}$ symmetry generated by
$\mathcal{R}=\mathcal{R}_{1}\mathcal{R}_{2}$. Hereafter we set $J_{\perp}^{x}=0$
and $J_{\perp}^{z}=1$, which describes a spatially isotropic transverse-field
Ising ladder supplemented by four-spin interactions. 

We are interested in the ground state (GS) phase diagram of the model
as a function of $B>0$ and $J_{4}<0$. To gain insight into the possible
phases, let us consider some particular limits of the Hamiltonian
parameters. For $J_{4}=0$, the physics is governed by the competition
between the ferromagnetic Ising coupling and the transverse magnetic
field. In this limit we expect a continuous phase transition in the
two-dimensional Ising universality class from a FM phase for $B<B_{c}$
to a paramagnetic (PM) phase for $B>B_{c}$. The FM phase has two
degenerate ground states that spontaneously break the $\mathcal{R}$
symmetry. In a four-site plaquette, the classical FM ground states
can be represented by $\begin{array}{cc}
+ & +\\
+ & +
\end{array}$ and $\begin{array}{cc}
- & -\\
- & -
\end{array}$, where $\pm$ denotes a spin polarized in the $\pm\hat{\mathbf{z}}$
direction. In the PM phase, the unique GS is adiabatically connected
with the product state with all spins polarized in the $\hat{\mathbf{x}}$
direction. On the other hand, for $B=0$ and $J_{4}\neq0$ we have
a classical Ising model with four-spin interactions. While $J_{4}>0$
favors the FM states, for $J_{4}<-3/2$ we find a new type of order
that we refer to as the dimerized-rung (DR) phase, which we will discuss
in the following. 

\begin{figure}
mv \includegraphics[scale=0.35]{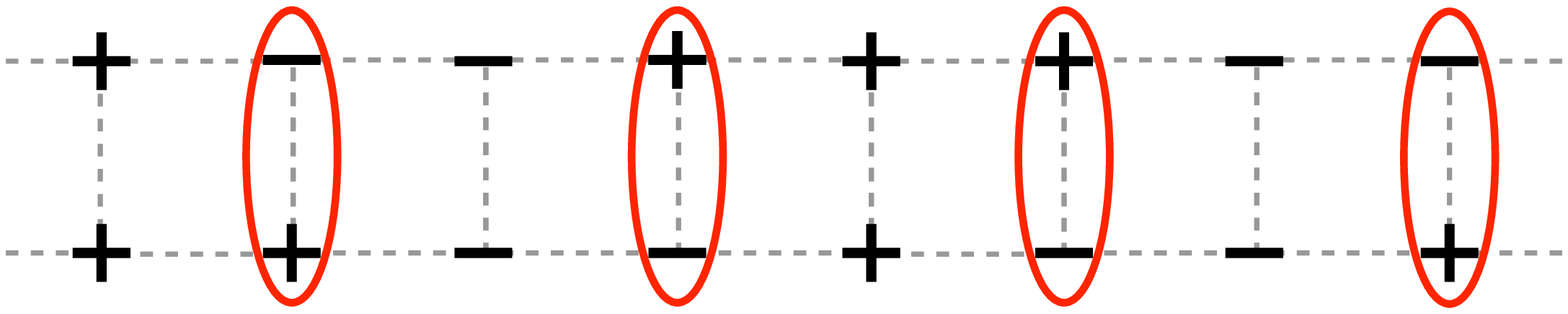}

\caption{\label{fig:drphase}Sketch of the dimerized-rung phase. The enclosed
up-down dimers are located in one sublattice.}
\end{figure}

The classical GSs of the DR phase obey the local constraint that all
plaquettes have only one spin with a different sign. An example is
sketched in Fig. \ref{fig:drphase}. Importantly, the constraint implies
an alternation between rungs with parallel or antiparallel spins.
Here it is convenient to introduce the pseudospins $T_{i}^{z}=\sigma_{1,i}^{z}\sigma_{2,i}^{z}$
and $S_{i}^{z}=\sigma_{1,i}^{z}$ and represent states in the local
basis for each rung by $\left|T_{i}^{z},S_{i}^{z}\right\rangle $.
The subspace of classical DR states with energy $E_{0}=-L|J_{4}|$
is then defined by $T_{i+1}^{z}=-T_{i}^{z}$ with arbitrary $\{S_{i}^{z}\}$.
As a consequence, the DR states break translational symmetry, $i\mapsto i+1$,
upon the choice of antiparallel spins, $T^{z}=-$, on the even or
odd sublattice. We refer to a rung with antiparallel spins as an up-down
dimer. The classical GS degeneracy is given by $2^{L+1}$. 

A small transverse magnetic field must lift the exponential degeneracy
of the classical model. To see this effect, we first rewrite the Hamiltonian
in terms of the rung pseudospins:
\begin{align}
{\cal H}= & -\sum_{i=1}^{L}\left[J_{4}T_{i}^{z}T_{i+1}^{z}+(1+T_{i}^{z}T_{i+1}^{z})S_{i}^{z}S_{i+1}^{z}+T_{i}^{z}\right]\nonumber \\
 & -B\sum_{i=1}^{L}(1+S_{i}^{x})T_{i}^{x},\label{Model2}
\end{align}
where $T_{i}^{x}=\sigma_{2,i}^{x}$ and $S_{i}^{x}=\sigma_{1,i}^{x}\sigma_{2,i}^{x}$,
so that the pseudospins obey $\{T_{i}^{x},T_{i}^{z}\}=\{S_{i}^{x},S_{i}^{z}\}=0$
and $[T_{i}^{x},S_{i}^{z}]=[S_{i}^{x},T_{i}^{z}]=0$. In this notation,
the global $\pi$ rotation is written as $\mathcal{R}=\prod_{i}S_{i}^{x}$.
The analysis of the Hamiltonian becomes particularly simple in the
limit $|J_{4}|,B\gg1$. Dropping the exchange interaction terms, we
obtain 
\begin{equation}
{\cal H}\approx-\sum_{i=1}^{L}\left[J_{4}T_{i}^{z}T_{i+1}^{z}+B(1+S_{i}^{x})T_{i}^{x}\right].\label{highBJ4}
\end{equation}
In this limit, the operators $S_{i}^{x}$ become conserved quantities.
If we fix $S_{i}^{x}=+1$ for all rungs, the model reduces to an Ising
chain with a uniform transverse field for the $T$ pseudospins. In
terms of the original spins, the state on each rung becomes a superposition:
\begin{align}
\left|T^{z}=+,S^{x}=+\right\rangle  & =\frac{1}{\sqrt{2}}\left(\left|\begin{array}{c}
+\\
+
\end{array}\right\rangle +\left|\begin{array}{c}
-\\
-
\end{array}\right\rangle \right),\\
\left|T^{z}=-,S^{x}=+\right\rangle  & =\frac{1}{\sqrt{2}}\left(\left|\begin{array}{c}
+\\
-
\end{array}\right\rangle +\left|\begin{array}{c}
-\\
+
\end{array}\right\rangle \right).
\end{align}

From now on we are going to consider only negative values of $J_{4}.$
For $B\ll-J_{4}$, the GSs are selected within the sector with $T_{i+1}^{z}=-T_{i}^{z}$.
The only remaining degeneracy is that associated with the broken translational
invariance. The two GSs can be viewed as crystals of up-down dimers
and are distinguished by the order parameter 
\begin{equation}
D=\frac{1}{L}\sum_{i=1}^{L}(-1)^{i}\left\langle T_{i}^{z}\right\rangle =\frac{1}{L}\sum_{i=1}^{L}(-1)^{i}\left\langle \sigma_{1,i}^{z}\sigma_{2,i}^{z}\right\rangle .\label{eq:D}
\end{equation}
Remarkably, the excited states with energy $\sim B\ll|J_{4}|$ also
originate from the sector with the local constraint $T_{i+1}^{z}=-T_{i}^{z}$;
thus, they all have $D\neq0$ in the thermodynamic limit. The elementary
excitation in the regime $B\ll|J_{4}|$ corresponds to inverting the
eigenvalue of $S_{i}^{x}$, by applying a phase flip to the rung state
\begin{equation}
\left|\begin{array}{c}
\sigma_{1}^{z}\\
\sigma_{2}^{z}
\end{array}\right\rangle \mapsto\sigma_{1}^{z}\left|\begin{array}{c}
\sigma_{1}^{z}\\
\sigma_{2}^{z}
\end{array}\right\rangle ,
\end{equation}
without disrupting the long-range DR order. 

As we increase the magnetic field in the regime $B,|J_{4}|\gg1$,
the quantum fluctuations suppress the DR order. According to Hamiltonian
(\ref{highBJ4}), there is an Ising transition at $B\approx|J_{4}|/2$
at which the DR order parameter vanishes. For $B\gg|J_{4}|\gg1$,
the GS has $T_{i}^{x}=S_{i}^{x}=+1$, which is equivalent to $\sigma_{1,i}^{x}=\sigma_{2,i}^{x}=+1$.
Thus, this high-field phase must be smoothly connected with the PM
phase found in the regime $B\gg1\gg|J_{4}|$. Note that once we restore
the interleg coupling $J_{\perp}^{z}=1$, corresponding to the last
term in the first line of Eq. (\ref{Model2}), the expectation value
$L^{-1}\sum_{i}\langle T_{i}^{z}\rangle$ becomes nonzero throughout
the entire phase diagram. We stress that the order parameter of the
DR phase is the staggered part of $\langle T_{i}^{z}\rangle$, whereas
the FM phase is characterized by $\langle S_{i}^{z}\rangle$, $\langle T_{i}^{z}S_{i}^{z}\rangle\neq0$.
While the above analysis predicts only DR order for $-J_{4}\gg1$,
nothing precludes the coexistence of DR and FM orders at intermediate
couplings. In fact, for the classical model with $J_{4}<-3/2$, the
subspace of degenerate GSs contains states with nonzero total magnetization
that satisfy the constraint $T_{i+1}^{z}=-T_{i}^{z}$. In Sec. \ref{sec:RESULTS}
we shall see that DR and FM orders coexist for $J_{4}\sim-1.5$ at
weak fields. 

\section{METHODOLOGY\label{sec:METHODOLOGY}}

As stated in the introduction, the purpose of this paper is to determine
the GS phase diagram of the model and to characterize the universality
classes of the critical lines using the entanglement entropy. Before
presenting our results, and for the sake of clarity, let us briefly
discuss how to infer the critical behavior from the scaling of the
Rényi entanglement entropies near second-order phase transitions.

Consider a one-dimensional system composed of subsystems ${\cal A}$
with $x$ sites ($x=1,\ldots,L$) and ${\cal B}$ with $L-x$ sites.
The $\alpha$-Rényi entanglement entropies (REE) of the GS are defined
as 

\begin{equation}
S_{\alpha}(L,x)=\frac{1}{1-\alpha}\ln\text{Tr}(\rho_{{\cal {A}}}^{\alpha}),\label{eq:renyientropy}
\end{equation}
where $\rho_{{\cal {A}}}=\mbox{Tr}_{{\cal {B}}}\rho$, constructed
from the GS, is the reduced density matrix of the subsystem ${\cal A}$.
The von Neumann entropy, also known as entanglement entropy, corresponds
to $\alpha\to1$.

The scaling of the REE of the GS is universal and we can explore this
universality to determine the critical behavior of the model. For
\emph{non-critical }systems\emph{,} the REE satisfies the entropic
area law (see Ref. \cite{RMP82-277} for a review). For one-dimensional
systems, the entropy is expected to approach a constant value at large
subsystem sizes, i.e., $S_{\alpha}(L,x\gg1)\rightarrow\ $$b_{\alpha}$.
On the other hand, for \emph{critical} systems with PBC, the REE is
expected to behave in the scaling regime $1\ll x\ll L$ as

\begin{equation}
S_{\alpha}^{\text{crit}}(L,x)=S_{\alpha}^{\text{CFT}}(L,x)+S_{\alpha}^{\text{USC}}(L,x).\label{eq:entropyb}
\end{equation}
The first term on the right-hand side of the above equation is the
leading correction predicted by the conformal field theory (CFT) and
is given by \citep{cold,cardyentan,entroreviewcalabrese,affleckboundary} 

\begin{equation}
S_{\alpha}^{\text{CFT}}=\frac{c}{6}\left(1+\frac{1}{\alpha}\right)\ln\left[\frac{L}{\pi}\sin\left(\frac{\pi x}{L}\right)\right]+a_{\alpha},\label{eq:entropyCFT}
\end{equation}
where $c$ is the central charge and $a_{\alpha}$ is a non-universal
constant. 

The unusual subleading correction, $S_{\alpha}^{\text{USC}}$, has
the following universal scaling \citep{entropyosc,xxPBCh,calabreseOBC,xavieralcarazosc,XavierAlca2012,cardyosc}
\begin{equation}
S_{\alpha}^{\text{USC}}=g_{\alpha}\cos(\kappa x+\phi)\left|\sin\left(\frac{\pi x}{L}\right)\right|^{-p_{\alpha}},\label{eq:entropyUnusual}
\end{equation}
where $g_{\alpha}$ is another non-universal constant and the exponent
$p_{\alpha}$ is related to the scaling dimension of the energy operator
$X_{\epsilon}$ by $p_{\alpha}=\frac{2X_{\epsilon}}{\alpha}$. The
wave vector $\kappa$ and the phase $\phi$ depend on the model. For
instance, $\kappa=0=\phi$ for the Ising model and $\kappa=\pi$,
$\phi=0$ for the spin-$s$ XXZ chains at zero magnetic field \citep{XavierAlca2012}.
For the XXX chain, the scaling is affected by a marginal operator
that accounts for logarithmic corrections. As a result, we must replace
the central charge in Eq. (\ref{eq:entropyCFT}) by \citep{cardyosc}

\begin{equation}
c_{\text{eff}}=c+\frac{1}{b^{2}}\left[\frac{2g}{1+\pi gb\ln\left[\frac{L}{\pi}\sin\left(\frac{\pi x}{L}\right)\right]}\right]^{3},\label{ceff}
\end{equation}
 where $g$ is a constant and $b$ is a universal coefficient in the
operator product expansion of the CFT. 

It is possible to explore the above scaling laws to determine the
critical behavior, as the procedure outlined in Ref. \citep{xavieralcarazQCP},
which we use here and explain in the following. For a fixed value
of the four-spin interaction, the pseudo-critical transverse field
$B_{c}^{L}$ is given by the maximum value of the entanglement entropy
difference (MVEED) defined by

\begin{equation}
\Delta S_{1}(L,B)=S_{1}(L,x=L/2)-S_{1}(L,x=L/4).\label{eq:diff}
\end{equation}
Thus, according to the different scalings of critical and non-critical
behavior presented above (see Ref. \citep{xavieralcarazQCP} for more
details), as $L\to\infty$, we should have 

\begin{equation}
\Delta S_{1}(L,B)=\begin{cases}
\frac{c}{6}\ln(2), & B=B_{c}^{L}\\
0, & B\ne B_{c}^{L}
\end{cases}.\label{eq:diff2}
\end{equation}
Note that one can use the above equation to get not only the critical
value of $B$, but also the finite-size estimate of the central charge,
given by $c^{L}=6\Delta S_{1}(L,B_{c}^{L})/\ln(2)$. Besides, we can
also estimate the dimension of the energy operator $X_{\epsilon}$
by fitting the numerical data with the following difference \citep{XavierAlca2012} 

\begin{equation}
d_{\alpha}(L,x)=S_{\alpha}(L,x)-\frac{c}{6}\left(1+\frac{1}{\alpha}\right)\ln\left[\frac{L}{\pi}\sin\left(\frac{\pi x}{L}\right)\right],\label{eq:diffentropy}
\end{equation}
which, asymptotically, behaves as

\begin{equation}
d_{\alpha}(L,x)=a_{\alpha}+g_{\alpha}\cos(\kappa\ell+\phi)\left|\sin\left(\frac{\pi x}{L}\right)\right|^{-p_{\alpha}}.\label{eq:diffentropyb}
\end{equation}

In order to compute the REE of the GS of the two-leg ladder model
described in Sec. \ref{sec:MODEL}, we have used the standard DMRG
approach keeping up to $m=400$ states per block in most of the case,
but for the multicritical point we kept up to $m=1000$ states. In
our DMRG procedure the center blocks are composed of 4 states, which
represent two sites. The discarded weight was typically $10^{-8}-10^{-12}$. 

\section{RESULTS\label{sec:RESULTS}}

We obtain the phase diagram by following the procedure described in
Sec. \ref{sec:METHODOLOGY} and computing the desired quantities for
different values of system size $L$. In what follows, we have considered
mostly lengths ranging from $L=8$ to $L=28$. In some instances,
such as when computing the energy exponent or the DR order parameter,
we have also used lengths up to $L=60$.

\subsection{Phase diagram and critical behavior}

First, we illustrate how accurate estimates of the critical points
$B_{c}(J_{4})$, as well of the central charge $c,$ can be obtained
using the procedure discussed above. Here we use PBC. Table I shows
a representative example of the finite-size estimates of $B_{c}^{L}$
and $c^{L}$ obtained from the maximum value of $\Delta S_{1}(L,B)$
for $J_{4}=0$. We clearly see that the transition from the FM to
the PM phase at $J_{4}=0$ has central charge $c=1/2$, and is therefore
in the same university class as the critical point of the transverse-field
Ising chain. 

\begin{table}
\begin{tabular}{ccc}
\toprule 
\multicolumn{3}{c}{\hspace*{0.5cm}$B_{c}^{L}$ at $J_{4}=0$}\tabularnewline
\midrule 
L & $B_{c}^{L}$ & $c^{L}$\tabularnewline
\midrule 
8 & 1.82660 & 0.50721 \tabularnewline
16 & 1.83171 & 0.50122\tabularnewline
20 & 1.83194 & 0.50077\tabularnewline
24 & 1.83213 & 0.50052\tabularnewline
28 & 1.83214 & 0.50038\tabularnewline
\midrule
extr. & 1.83214(5) & \tabularnewline
\bottomrule
\end{tabular}

\caption{Finite-size estimates of $B_{c}^{L}$ and $c^{L}$ for $J_{4}=0$
obtained from the MVEED method; see Eqs. (\ref{eq:diff}) and (\ref{eq:diff2}).}
\end{table}

\begin{table}
\begin{tabular}{cccc}
\hline 
L & $J_{4}^{\scriptsize\text{mc}}$ & $B^{\scriptsize\text{mc}}$ & $c^{L}$\tabularnewline
\hline 
8 & 1.9381 & 0.5925 & 1.1451\tabularnewline
12 & 1.9395 & 0.5753 & 1.0701\tabularnewline
16 & 1.9446 & 0.5877 & 1.0504\tabularnewline
20 & 1.9450 & 0.5863 & 1.0214\tabularnewline
24 & 1.9452 & 0.5879 & 1.0132\tabularnewline
28 & 1.9456 & 0.5875 & 1.0109\tabularnewline
\hline 
\end{tabular}

\caption{Finite-size estimates of $B^{\scriptsize\text{mc}}$ and $J_{4}^{\scriptsize\text{mc}}$,
and $c^{L}$ for the multicritical point.}
\end{table}

\begin{figure}
\includegraphics[scale=0.35]{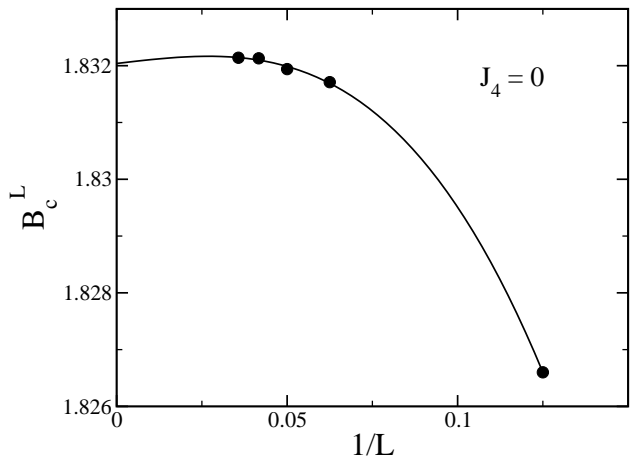}

\caption{\label{fig:BxL}Critical transverse field $B_{c}^{L}$ for $J_{4}=0$,
estimated from the maximum value of the entanglement entropy, as a
function of $1/L$. The solid line is a fit to Eq. (\ref{eq:Bcscalin}). }
\end{figure}

The finite-size estimates for the critical field for $J_{4}=0$ in
Table I are consistent with the value $B_{c}=1.838$ found in Ref.
\citep{arealawflavia}. The estimate in Ref. \citep{arealawflavia}
was determined after extrapolating the data for a two-leg ladder with
open boundary conditions to the thermodynamic limit $L\to\infty$.
A similar extrapolation is performed in Fig. \ref{fig:BxL}, where
we show the results for the critical field from Table I, as a function
of the inverse length, fitted according to the scaling equation
\begin{equation}
B_{c}^{L}=B_{c}+aL^{-1/\nu}(1+bL^{-\omega}),\label{eq:Bcscalin}
\end{equation}
where $B_{c}$ is the critical field in the thermodynamic limit, $a$
and $b$ are non-universal constants, and $\nu=1$ and $\omega=2$
are the correlation length and correction-to-scaling Ising critical
exponents, respectively. We obtain the extrapolated value $B_{c}=1.83214(5)$,
which we believe is more accurate than the previous value from Ref.
\citep{arealawflavia}. 

\begin{figure}[tp]
\begin{centering}
\includegraphics[scale=0.35]{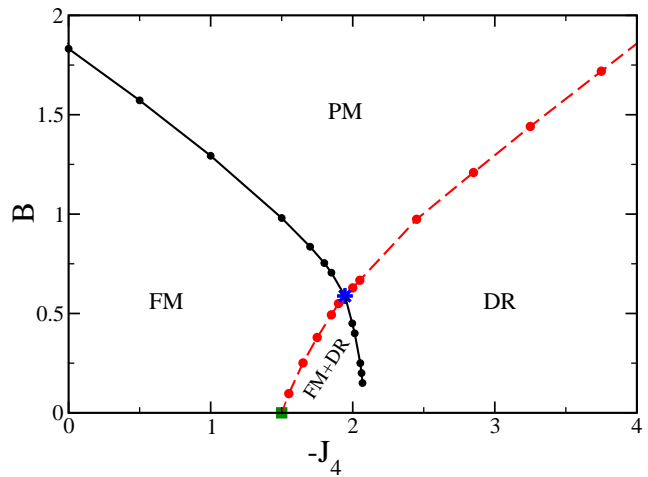}
\par\end{centering}
\caption{\label{pd} Phase diagram of the two-leg ladder as a function of the
\emph{negative} of the four-spin interaction $J_{4}$ and transverse
field $B$, showing the ferromagnetic (FM), dimerized-rung (DR), and
paramagnetic (PM) phases. The circles are the numerical finite-size
estimates of the critical points for system size $L=28$. The solid
and dashed lines correspond to Ising transitions. The square at $B=0$,
$J_{4}=-3/2$ corresponds to a classical transition point and the
star, at the crossing of the critical lines, to a multicritical point.
Below the multicritical point there is a region of coexistence of
FM and DR order parameters. }
\end{figure}

We have also determined $B_{c}^{L}$ and $c^{L}$ for several other
values of $J_{4}$ and system sizes $L$. Similarly to the results
for $J_{4}=0$ in Table I, the critical fields for $L=28$ are already
in excellent agreement with the extrapolation to the thermodynamic
limit. For this reason, in Fig. \ref{pd} we present the critical
lines estimated from data for $L=28$. 

Our numerical results in Fig. \ref{pd} reveal the presence of two
critical lines, both with central charge $c=1/2$, which cross at
a multicritical point. As discussed in Sec. \ref{sec:MODEL}, at large
$|J_{4}|$ the critical line corresponds to the Ising transition from
the DR to the PM phase. Thus, dashed line in Fig. \ref{pd} can be
associated with the spontaneous breaking of lattice translational
symmetry. In constrast, the solid line represents the Ising transition
at which the spin-rotation $\mathcal{R}$ symmetry is broken. By this
reasoning, we expect that in the region below the multicritical point
both symmetries are broken, leading to a coexistence of the FM and
DR phases. We will confirm this picture by computing the order parameters
in subsection \ref{subsec:Order-parameters}. In the following we
proceed with characterizing the multicritical point. 

The finite-size estimates of $c^{L}$ indicate that the central charge
is larger at the crossing of the two Ising transition lines. We locate
the multicritical point ($B^{\scriptsize\text{mc}},J_{4}^{\scriptsize\text{mc}}$)
precisely by finding the maximum value of $\Delta S_{L}(B,J_{4})$
in the $J_{4}$-$B$ plane. The same procedure was used in Ref. \citep{xavieralcarazQCP}
to determine the tricritical point of the Blume-Capel model. In Table
II, we present the finite-size estimates of the multicritical couplings
obtained by this procedure, together with the estimates of the central
charge $c^{L}.$ We clearly see that the universality class of the
multicritical point is described by a CFT with $c=1$.

Phenomenologically, the central charge can be understood in terms
of two independent Majorana fermions which become massless at the
Ising transitions represented by the lines in Fig. \ref{pd}. Since
these transitions break different symmetries, there is no direct coupling
between the order parameters of these two Ising CFTs. At the multicritical
point, the two species of massless Majorana fermions can be combined
to define a complex fermion, which can then be bosonized \citep{Gogolin}.
Thus, this point belong to the Gaussian universality class, and the
effective field theory with $c=1$ can be written in terms of a single
massless boson. This theory allows for local operators whose scaling
dimension varies continuosly as a function of a Luttinger parameter
$K=K(B^{\scriptsize\text{mc}},J_{4}^{\scriptsize\text{mc}})$. The
effective field theory for the multicritical point is discussed in
more detail in Appendix A. 

\begin{figure}[tp]
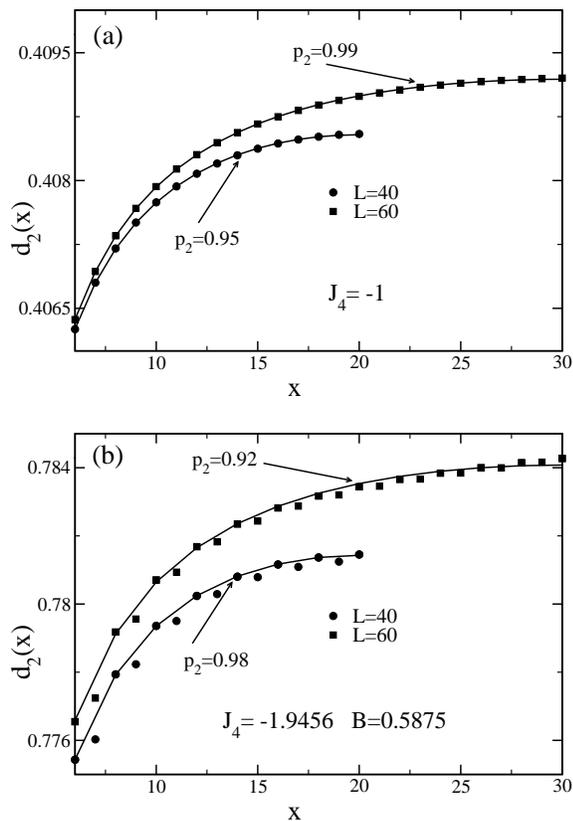

\begin{centering}
\includegraphics[scale=0.3]{fig4a}
\par\end{centering}
\vspace*{\bigskipamount}

\begin{centering}
\includegraphics[scale=0.3]{fig4b}
\par\end{centering}
\vspace*{\bigskipamount}

\caption{\label{fig2} Results for $d_{2}(L,x)$ versus $x$ for the two-leg
ladder model with PBC. (a) Data for $J_{4}=-1$ and two values of
$L$ (see legend). (b) Same as (a), but for the multicritical point.
The symbols are the numerical data and the solid lines connect the
fitted points using Eq. (\ref{eq:diffentropyb}) with $c=1/2$ in
(a) and $c=1$ in (b). The arrows indicate the values of $p_{2}\equiv X_{\epsilon}$
obtained through the fit. We discarded the first 4 points of $d_{2}$
in the fit procedure.}
\end{figure}

To strengthen the case for the crossing of two Ising transitions,
we determine the scaling dimension $X_{\epsilon}$ of the energy operator
using the fitting procedure summarized by Eqs. (\ref{eq:diffentropy})
and (\ref{eq:diffentropyb}). In Fig. \ref{fig2}(a), we present a
representative example of the difference $d_{2}(L,x)$ as a function
of $x$ for $J_{4}=-1$ and two system sizes: $L=40$ and $L=60$.
We performed similar fits along the entire critical lines and obtained
the exponent $p_{2}=X_{\epsilon}\approx1$. These results are consistent
with the fact the universality class of critical behavior of these
critical lines is the same as the quantum Ising chain, where $c=1/2$
and $X_{\epsilon}=1$. 

We also estimate $X_{\epsilon}$ at the multicritical point where
$c=1.$ It is important to mention here that, for large system sizes,
it is necessary to use more states per block in the DMRG procedure.
For this reason, at the multcritical point we use $m=1000$ states
per block. In Fig. \ref{fig2}(b), we show the difference $d_{2}(L,x)$
at the multicritical point. Note that in this case the subleading
corrections exhibit oscillations. Similar oscillations are routinely
observed in the entanglement entropy of critical systems described
by $c\geq1$ CFTs and present tendency of antiferromagnetic order,
for instance the spin-$1/2$ XXZ chain \citep{xavieralcarazosc}.
In that case, the gapless bosonic mode of the $c=1$ CFT arises naturally
from the U(1) symmetry of the model. By contrast, in our model the
U(1) symmetry of the low-energy fixed point is emergent and requires
fine tuning to the multicritical point. In our present case, it seems
that the origin of the oscillations comes from the dimerization, which
also alternate the spins in the lattice in analogy to the antiferromagnetic
order. To remove the effect of the oscillations and make the analysis
easier, we fit the numerical data using Eq. (\ref{eq:diffentropyb})
considering only $x$ even. Our results indicate that $X_{\epsilon}\approx1$
at the multicritical point. 

\subsection{Order parameters\label{subsec:Order-parameters}}

We close this section by presenting results for the order parameter
associated with the DR and FM phases. For practical purposes, instead
of using Eq. (\ref{eq:D}), we consider the parameter defined locally
on a four-site plaquette 
\begin{equation}
D_{i}=\frac{1}{2}|\langle\sigma_{1,i+1}^{z}\sigma_{2,i+1}^{z}-\sigma_{1,i}^{z}\sigma_{2,i}^{z}\rangle|.\label{eq:drop}
\end{equation}
Clearly, in the translationally-invariant GSs for the FM and PM phases
we have $D_{i}=0$ for any plaquette. Inside the DR phase, however,
we must be careful because $D_{i}=0$ for any \emph{finite }system
with PBC. In order to investigate the symmetry breaking by considering
finite\emph{ }systems, we add a small perturbation that selects one
of the two DR GSs which become exactly degenerate in the thermodynamic
limit. A simple way to lift this degeneracy is to consider a system
with semi-open boundary conditions (SOBC), consisting of an open ladder
with $L$ odd to which we add the following term that connects the
two edges: 

\begin{equation}
H_{\text{BC}}^{\text{DR}}=-\left(\sigma_{1,1}^{z}\sigma_{1,L}^{z}+\sigma_{2,1}^{z}\sigma_{2,L}^{z}\right).\label{eq:SOBC}
\end{equation}
Similarly, we can view this SOBC as the ladder with PBC in which we
supress the the four-spin interaction of the $L$-th plaquette. We
then consider the order parameter at the center of the ladder, $D_{(L+1)/2}$,
far from the perturbation at the boundary. 

\begin{figure}[H]
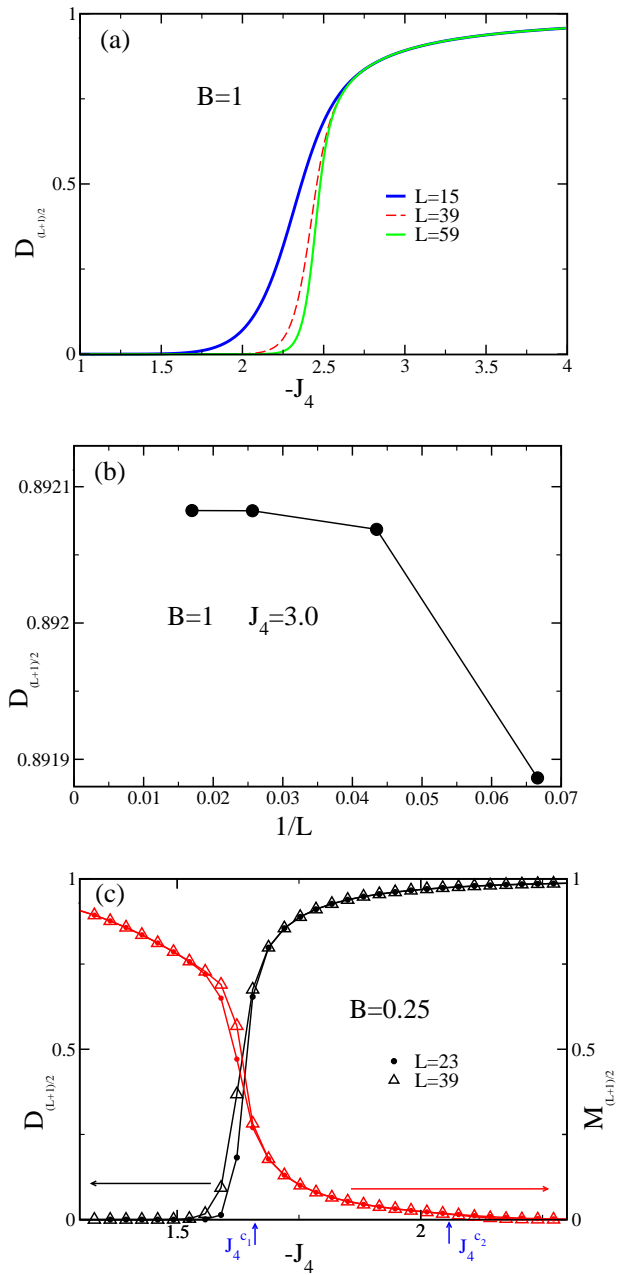

\begin{centering}
\includegraphics[scale=0.3]{fig5a}
\par\end{centering}
\vspace*{\bigskipamount}

\begin{centering}
\includegraphics[scale=0.3]{fig5b}
\par\end{centering}
\vspace*{\bigskipamount}

\begin{centering}
\hspace*{0.7cm}\includegraphics[scale=0.3]{fig5c}
\par\end{centering}
\caption{\label{fig3} Order parameters for the phases of the two-leg ladder
with a particular boundary condition, see Eq. (\ref{eq:SOBC}). (a)
DR order paramaeter as a function of $J_{4}$ for $B=1$ and different
system sizes $L$. (b) Finite size scaling of the DR order parameter
for $J_{4}=-3$ and $B=1$. (c) $D_{(L+1)/2}$ and $M_{(L+1)/2}$
versus $J_{4}$ for $B=0.25$ and two values of $L$. The blue arrows
in the horizotal axis indicate the positions of the two critical points
for $B=0.25$ according to the phase diagram in Fig. \ref{pd}. The
lines in (b) and (c) connect the numerical data.}
\end{figure}

To illustrate the behavior of the DR order parameter, we first present
in Fig. \ref{fig3}(a) the result for $D_{(L+1)/2}$ as a function
of $J_{4}$ for $B=1$ and different values of system size $L$. Note
that, according to the phase diagram in Fig. \ref{pd}, as we vary
$J_{4}$ at fixed $B=1$ we cut accross three different phases. As
expected, $D_{(L+1)/2}$ is nonzero only in the DR phase. The finite
size scaling analysis in Fig. \ref{fig3}(b) confirms that $D_{(L+1)/2}$
approaches a finite value at large values $|J_{4}|$, providing unambiguous
evidence of the DR GS. 

We can also locate the transition corresponding to the dashed line in
Fig. \ref{pd} by analyzing the DR order parameter. For a fixed value
of $B$, we can estimate the critical point $J_{4}^{c}$ as the inflection
point of $D_{(L+1)/2}$ as a function of $J_{4}$. We find that the
estimates of the critical points obtained within this procedure (not
shown) are in good agreement with the results from the MVEED. 

Finally, we investigate the FM order parameter defined as $M_{(L+1)/2}=\langle\sigma_{1,(L+1)/2}^{z}\rangle$.
Similarly to the DR order parameter, $M_{(L+1)/2}$ vanishes identically
for any\emph{ }finite system with PBC. We must then add a local perturbation
that breaks the $\mathbb{Z}_{2}$ spin-rotation symmetry. We consider
a weak boundary longitudinal field that scales with system size:

\begin{equation}
H_{\text{BC}}^{\text{FM}}=-\frac{1}{L^{2}}\left(\sigma_{1,1}^{z}+\sigma_{2,1}^{z}\right).
\end{equation}
In Fig. \ref{fig3}(c) we show the results for $D_{(L+1)/2}$ and
$M_{(L+1)/2}$ as a function of $J_{4}$ for $B=0.25$ and two different
system sizes. For this value of $B$ we have two critical points,
$J_{4}^{c_{1}}\approx-1.65$ and $J_{4}^{c_{2}}\approx-2.05$, which
are indicated by blue arrows in the horizontal axis of Fig. \ref{fig3}(c).
As we observe in this figure, both order parameters are nonzero between
the two critical points, demonstrating that the FM and DR orders coexist
in this range. Finite size effects are more pronounced for $J_{4}\approx J_{4}^{c_{1}}$
. We have performed a finite size scaling analysis similar to the
one in Fig. \ref{fig3}(b) and verified that $\lim_{L\rightarrow\infty}M_{(L+1)/2}$
is nonzero for some values of $J_{4}$ above $J_{4}^{c_{1}}$ and
below $J_{4}^{c_{2}}$. 

For $B\ll1$, the onset of DR order matches the classical transition
point $J_{4}=-3/2$, see Fig. \ref{pd}. As discussed in Sec. \ref{sec:MODEL},
the classical model with $B=0$ has an exponentially degenerate GS
manifold which includes DR states with finite magnetization. Our observation
of a coexistence phase implies that, slightly above the classical
transition point, an arbitrarily small transverse field lifts the
exponential degeneracy and selects four GSs that break both spin-rotation
and translational symmetries. These four GSs can be labeled by the
signs of $D_{(L+1)/2}$ and $M_{(L+1)/2}$ in the limit $L\to\infty$.
In addition, for $B>0$ there appears the second critical point, beyond
which the FM order parameter vanishes and the spin-rotation symmetry
is restored, while the translational symmetry remains broken. Using
the MVEED, we have not been able to track the transition from the
coexistence phase to the DR phase all the way down to $B\to0$; see
the black line in Fig. \ref{pd}. The numerical difficulties for small
$B$ and\textcolor{red}{{} }$-J_{4}>3/2$ are most likely associated
with the approximate degeneracy of exponentially many states in this
regime. 

\section{CONCLUSIONS\label{sec:CONCLUSIONS}}

We investigated a transverse-field two-leg Ising ladder with a plaquette
four-spin interaction. We obtained the ground state phase diagram
by exploring the finite-size scaling entanglement entropy. We found
two critical lines that separate two ordered phases (see Fig. \ref{pd}):
a ferromagnetic and a dimerized-rung phase. These phases spontaneously
break the symmetries of $\mathbb{Z}_{2}$ spin rotation and translation
by one site, respectively. The universality class of both transitions
is the same as the quantum Ising chain. We confirmed this critical
behavior by calculating the central charge using the entanglement
entropy. We found that the critical lines with central charge $c=1/2$
cross at a multicritical point with $c=1$ (see Table II). By analyzing
the subleading corrections of the scaling Rényi entanglement entropy,
we were able to determine the scaling dimension $X_{\epsilon}$ of
the energy operator. Our results support $X_{\epsilon}=1$ along the
critical lines, as expected for models with $c=1/2,$ as well as at
the multicritical point. By investigating the order parameters, we
found a region in parameter space where the ferromagnetic and dimerized-rung
orders coexist (see Fig. \ref{fig3}). It is interesting to notice
that the plaquette symmetry of the four-spin interaction seems to
be responsible for the richness of the phase diagram presented by
this model. In contrast, the transverse-field chain in the regime
of dominant linear four-spin interaction has only an eightfold ground
state degeneracy and a first-order transition to the disordered state
takes place instead, with the multicritical point located at zero
transverse field \citep{ozi2}.
\begin{acknowledgments}
We would like to thank Prof. F. C. Alcaraz for fruitfull discussions
and Prof. G. Weber for the use of the workstations of the Statistical
Physics group at UFMG in Brazil. We also thank the resources and technical
expertise from the Georgia Advanced Computing Resource Center (GACRC),
a partnership between the University of Georgia's Office of the Vice
President for Research and Office of the Vice President for Information
Technology. We would like to address invaluable help from Shan-Ho
Tsai regarding the use of the GACRC computing facilities at UGA. This
work was supported by the Brazilian agencies CNPq, CAPES and FAPEMIG.
Research at IIP-UFRN is supported by Brazilian ministries MEC and
MCTI and by a grant from Associação Instituto Internacional de Física.
\end{acknowledgments}

\appendix

\section{Effective field theory for the multicritical point}

In this appendix we analyze the effective field theory that describes
the vicinity of the multicritical point with $c=1$ in the two-leg
ladder model. Since the multicritical point appears at intermediate
couplings, far from any exactly solvable point in the phase diagram,
we adopt a phenomenological approach guided by symmetry considerations.
We can cast the order parameters of the model in terms of two $\mathbb{Z}_{2}$
symmetries: the spin-rotation symmetry $\mathcal{R}$ and the symmetry
of exchanging even and odd sublattices upon translation by one site.
Let $\sigma_{1}(x)$ and $\sigma_{2}(x)$ denote the order operators
in the Ising CFTs \citep{critical3} that describe the low-energy
physics near the transitions where these $\mathbb{Z}_{2}$ symmetries
are spontaneously broken. The order operators have conformal weights
$(\frac{1}{16},\frac{1}{16})$. The FM phase corresponds to $\langle\sigma_{1}\rangle\neq0$,
$\langle\sigma_{2}\rangle=0$, whereas the DR phase corresponds to
$\langle\sigma_{1}\rangle=0$, $\langle\sigma_{2}\rangle\neq0$. In
the coexistence phase, we have $\langle\sigma_{1}\rangle\neq0$ and
$\langle\sigma_{2}\rangle\neq0$. 

The effective Hamiltonian density for the perturbed CFT including
all local operators allowed by symmetry is 
\begin{align*}
\mathcal{H}_{\text{eff}} & =\sum_{n=1,2}\left[\frac{iv_{n}}{2}\left(\xi_{n}\partial_{x}\xi_{n}-\bar{\xi}_{n}\partial_{x}\bar{\xi}_{n}\right)+im_{n}\xi_{n}\bar{\xi}_{n}\right]\\
 & \qquad+g\xi_{1}\bar{\xi}_{1}\xi_{2}\bar{\xi}_{2},
\end{align*}
where $\xi_{n}(x)$ and $\bar{\xi}_{n}(x)$ are chiral Majorana fermions
with conformal weights $(\frac{1}{2},0)$ and $(0,\frac{1}{2})$,
respectively, which obey $\{\xi_{n}(x),\xi_{n'}(x')\}=\{\bar{\xi}_{n}(x),\bar{\xi}_{n'}(x')\}=\delta_{nn'}\delta(x-x')$
and $\{\xi_{n}(x),\bar{\xi}_{n'}(x')\}=0$. The operators $\epsilon_{n}=i\xi_{n}\bar{\xi}_{n}$,
with conformal weights $(\frac{1}{2},\frac{1}{2})$ and scaling dimension
$X_{\epsilon}=1$, can be identified with the energy operators in
the Ising CFTs. The velocities $v_{1}$ and $v_{2}$ can be different
since the two species of Majorana fermions are not related by symmetry.
The masses $m_{1}$ and $m_{2}$ can also be different and represent
the couplings constants for the leading relevant operators in the
effective Hamiltonian. Importantly, the term $\sigma_{1}\sigma_{2}$,
which would be a strongly relevant perturbation, is not allowed because
the order parameters change sign under different $\mathbb{Z}_{2}$
symmetries. As a result, the leading interaction between the two Ising
models is the quartic interaction with coupling constant $g$. 

The CFT with a single gapless Majorana fermion has central charge
$c=1/2$. Thus, the critical lines in the phase diagram in Fig. \ref{pd}
are defined by the condition that one of the masses goes through zero
while the other one remains finite. At the crossing of these lines,
we have $m_{1}=m_{2}=0$. At this point the CFT of two decoupled Ising
models is only perturbed by the quartic interaction with coupling
constant $g$. To analyze the effects of this interaction, we define
a complex fermion from the linear combination of the Majorana fermions:
\[
\psi=\frac{\xi_{1}+i\xi_{2}}{\sqrt{2}},\qquad\bar{\psi}=\frac{\bar{\xi}_{1}+i\bar{\xi}_{2}}{\sqrt{2}}.
\]
We can then bosonize the chiral complex fermions in the form $\psi\sim e^{-i\sqrt{\pi}(\theta+\phi)}$,
$\bar{\psi}\sim e^{-i\sqrt{\pi}(\theta-\phi)}$, where $\phi(x)$
and $\theta(x)$ are dual bosonic fields that obey $[\phi(x),\partial_{x'}\theta(x')]=i\delta(x-x')$
\citep{Gogolin}. The bosonized Hamiltonian density for $m_{1}=m_{2}=0$
has the form 
\begin{align*}
\mathcal{H}_{\text{eff}}^{\text{mc}} & =\frac{vK}{2}(\partial_{x}\theta)^{2}+\frac{v}{2K}(\partial_{x}\phi)^{2}+\lambda_{1}\cos(4\sqrt{\pi}\phi)\\
 & \qquad+\lambda_{2}\cos(4\sqrt{\pi}\theta)+\lambda_{3}\cos(\sqrt{4\pi}\theta)\cos(\sqrt{4\pi}\phi),
\end{align*}
where $v$ is the renormalized velocity, $K$ is the Luttinger parameter,
and the cosine terms have coupling constants $\lambda_{1},\lambda_{2}\sim g$
and $\lambda_{3}\sim v_{1}-v_{2}$. For $g=0$, we have $K=1$, corresponding
to free fermions. The $\lambda_{3}$ term, associated with the velocity
mismatch, has scaling dimension $K+K^{-1}$ and is irrelevant for
any $K\neq1$. The $\lambda_{1}$ and $\lambda_{2}$ terms have scaling
dimensions $4K$ and $4/K$, respectively. As a result, they are both
irrelevant for a wide range of the Luttinger parameter, $1/2<K<2$.
Dropping the irrelevant operators, we conclude that the low-energy
fixed point at the crossing of the critical lines corresponds to a
free boson, with central charge $c=1$. The scaling dimensions of
local operators depend on the value of the Luttinger parameter, which
is not fixed by any symmetries. 

\bibliographystyle{apsrev4-1}
\bibliography{refs_rev4}

\begin{thebibliography}{63}%
\makeatletter
\providecommand \@ifxundefined [1]{%
 \@ifx{#1\undefined}
}%
\providecommand \@ifnum [1]{%
 \ifnum #1\expandafter \@firstoftwo
 \else \expandafter \@secondoftwo
 \fi
}%
\providecommand \@ifx [1]{%
 \ifx #1\expandafter \@firstoftwo
 \else \expandafter \@secondoftwo
 \fi
}%
\providecommand \natexlab [1]{#1}%
\providecommand \enquote  [1]{``#1''}%
\providecommand \bibnamefont  [1]{#1}%
\providecommand \bibfnamefont [1]{#1}%
\providecommand \citenamefont [1]{#1}%
\providecommand \href@noop [0]{\@secondoftwo}%
\providecommand \href [0]{\begingroup \@sanitize@url \@href}%
\providecommand \@href[1]{\@@startlink{#1}\@@href}%
\providecommand \@@href[1]{\endgroup#1\@@endlink}%
\providecommand \@sanitize@url [0]{\catcode `\\12\catcode `\$12\catcode
  `\&12\catcode `\#12\catcode `\^12\catcode `\_12\catcode `\%12\relax}%
\providecommand \@@startlink[1]{}%
\providecommand \@@endlink[0]{}%
\providecommand \url  [0]{\begingroup\@sanitize@url \@url }%
\providecommand \@url [1]{\endgroup\@href {#1}{\urlprefix }}%
\providecommand \urlprefix  [0]{URL }%
\providecommand \Eprint [0]{\href }%
\providecommand \doibase [0]{http://dx.doi.org/}%
\providecommand \selectlanguage [0]{\@gobble}%
\providecommand \bibinfo  [0]{\@secondoftwo}%
\providecommand \bibfield  [0]{\@secondoftwo}%
\providecommand \translation [1]{[#1]}%
\providecommand \BibitemOpen [0]{}%
\providecommand \bibitemStop [0]{}%
\providecommand \bibitemNoStop [0]{.\EOS\space}%
\providecommand \EOS [0]{\spacefactor3000\relax}%
\providecommand \BibitemShut  [1]{\csname bibitem#1\endcsname}%
\let\auto@bib@innerbib\@empty
\bibitem [{\citenamefont {{J. B. Parkinson}}\ and\ \citenamefont {{D. J. J.
  Farnell}}(2010)}]{park}%
  \BibitemOpen
  \bibfield  {author} {\bibinfo {author} {\bibnamefont {{J. B. Parkinson}}}\
  and\ \bibinfo {author} {\bibnamefont {{D. J. J. Farnell}}},\ }\href@noop {}
  {\emph {\bibinfo {title} {An {I}ntroduction to {Q}uantum {S}pin {S}ystems}}}\
  (\bibinfo  {publisher} {Springer, Berlin Heidelberg},\ \bibinfo {year}
  {2010})\ \bibinfo {note} {{L}ect. Notes Phys. 816}\BibitemShut {NoStop}%
\bibitem [{\citenamefont {{W. D. Ratcliff}}\ and\ \citenamefont {{W.
  Lynn}}(2015)}]{rat}%
  \BibitemOpen
  \bibfield  {author} {\bibinfo {author} {\bibnamefont {{W. D. Ratcliff}}}\
  and\ \bibinfo {author} {\bibnamefont {{W. Lynn}}},\ }\href@noop {} {\emph
  {\bibinfo {title} {Neutron {S}cattering - {M}agnetic and {Q}uantum
  {P}henomena, {E}xperimental {M}ethods in the {P}hysical {S}ciences}}},\
  edited by\ \bibinfo {editor} {\bibfnamefont {F.}~\bibnamefont
  {Fernandez-Alonso}}\ and\ \bibinfo {editor} {\bibfnamefont {D.~L.}\
  \bibnamefont {Price}},\ Vol.~\bibinfo {volume} {48}\ (\bibinfo  {publisher}
  {Elsevier, Cambridge},\ \bibinfo {year} {2015})\BibitemShut {NoStop}%
\bibitem [{\citenamefont {{A. L. Patrick}}\ \emph {et~al.}(2006)\citenamefont
  {{A. L. Patrick}}, \citenamefont {{N. Nagaosa}},\ and\ \citenamefont {{X-G
  Wen}}}]{RevModPhys.78.17}%
  \BibitemOpen
  \bibfield  {author} {\bibinfo {author} {\bibnamefont {{A. L. Patrick}}},
  \bibinfo {author} {\bibnamefont {{N. Nagaosa}}}, \ and\ \bibinfo {author}
  {\bibnamefont {{X-G Wen}}},\ }\href@noop {} {\bibfield  {journal} {\bibinfo
  {journal} {Rev. Mod. Phys.}\ }\textbf {\bibinfo {volume} {78}},\ \bibinfo
  {pages} {17} (\bibinfo {year} {2006})}\BibitemShut {NoStop}%
\bibitem [{\citenamefont {{J. Cho}}\ \emph {et~al.}(2004)\citenamefont {{J.
  Cho}}, \citenamefont {{Y. Fujii}}, \citenamefont {{K. Konioshi}},
  \citenamefont {{J. Yoon}}, \citenamefont {{N. Kim}}, \citenamefont {{J.
  Jung}}, \citenamefont {{S. Miwa}}, \citenamefont {{M. Jung}}, \citenamefont
  {{Y. Suzuki}},\ and\ \citenamefont {{C. You}}}]{cho}%
  \BibitemOpen
  \bibfield  {author} {\bibinfo {author} {\bibnamefont {{J. Cho}}}, \bibinfo
  {author} {\bibnamefont {{Y. Fujii}}}, \bibinfo {author} {\bibnamefont {{K.
  Konioshi}}}, \bibinfo {author} {\bibnamefont {{J. Yoon}}}, \bibinfo {author}
  {\bibnamefont {{N. Kim}}}, \bibinfo {author} {\bibnamefont {{J. Jung}}},
  \bibinfo {author} {\bibnamefont {{S. Miwa}}}, \bibinfo {author} {\bibnamefont
  {{M. Jung}}}, \bibinfo {author} {\bibnamefont {{Y. Suzuki}}}, \ and\ \bibinfo
  {author} {\bibnamefont {{C. You}}},\ }\href@noop {} {\bibfield  {journal}
  {\bibinfo  {journal} {J. Mag. Mag. Mat.}\ }\textbf {\bibinfo {volume} {76}},\
  \bibinfo {pages} {323} (\bibinfo {year} {2004})}\BibitemShut {NoStop}%
\bibitem [{\citenamefont {{ I. \ifmmode \check{Z}\else \v{Z}\fi{}uti\ifmmode
  \acute{c}\else\'{c}\fi{}}}\ \emph {et~al.}(2004)\citenamefont {{ I. \ifmmode
  \check{Z}\else \v{Z}\fi{}uti\ifmmode \acute{c}\else\'{c}\fi{}}},
  \citenamefont {{F. Jaroslav}},\ and\ \citenamefont {{S. Das
  Sarma}}}]{RevModPhys.76.323}%
  \BibitemOpen
  \bibfield  {author} {\bibinfo {author} {\bibnamefont {{ I. \ifmmode
  \check{Z}\else \v{Z}\fi{}uti\ifmmode \acute{c}\else\'{c}\fi{}}}}, \bibinfo
  {author} {\bibnamefont {{F. Jaroslav}}}, \ and\ \bibinfo {author}
  {\bibnamefont {{S. Das Sarma}}},\ }\href@noop {} {\bibfield  {journal}
  {\bibinfo  {journal} {Rev. Mod. Phys.}\ }\textbf {\bibinfo {volume} {76}},\
  \bibinfo {pages} {323} (\bibinfo {year} {2004})}\BibitemShut {NoStop}%
\bibitem [{\citenamefont {{J. L. Jimenez}}\ \emph {et~al.}(2021)\citenamefont
  {{J. L. Jimenez}}, \citenamefont {{ S. P. G. Crone}},\ and\ \citenamefont
  {{E. Fogh et al.}}}]{jim}%
  \BibitemOpen
  \bibfield  {author} {\bibinfo {author} {\bibnamefont {{J. L. Jimenez}}},
  \bibinfo {author} {\bibnamefont {{ S. P. G. Crone}}}, \ and\ \bibinfo
  {author} {\bibnamefont {{E. Fogh et al.}}},\ }\href@noop {} {\bibfield
  {journal} {\bibinfo  {journal} {Nature}\ }\textbf {\bibinfo {volume} {592}},\
  \bibinfo {pages} {370} (\bibinfo {year} {2021})}\BibitemShut {NoStop}%
\bibitem [{\citenamefont {{P. R. C. Guimaraes}}\ \emph
  {et~al.}(2015)\citenamefont {{P. R. C. Guimaraes}}, \citenamefont {{J. A.
  Plascak}}, \citenamefont {{O. F. de Alcantara Bonfim}},\ and\ \citenamefont
  {{J. Florencio}}}]{cola}%
  \BibitemOpen
  \bibfield  {author} {\bibinfo {author} {\bibnamefont {{P. R. C. Guimaraes}}},
  \bibinfo {author} {\bibnamefont {{J. A. Plascak}}}, \bibinfo {author}
  {\bibnamefont {{O. F. de Alcantara Bonfim}}}, \ and\ \bibinfo {author}
  {\bibnamefont {{J. Florencio}}},\ }\href@noop {} {\bibfield  {journal}
  {\bibinfo  {journal} {Phys. Rev. E}\ }\textbf {\bibinfo {volume} {92}},\
  \bibinfo {pages} {042115} (\bibinfo {year} {2015})}\BibitemShut {NoStop}%
\bibitem [{\citenamefont {{B. Boechat}}\ \emph {et~al.}(2014)\citenamefont {{B.
  Boechat}}, \citenamefont {{J. Florencio}}, \citenamefont {{A. Saguia}},\ and\
  \citenamefont {{O. F. de Alcantara Bonfim}}}]{bia}%
  \BibitemOpen
  \bibfield  {author} {\bibinfo {author} {\bibnamefont {{B. Boechat}}},
  \bibinfo {author} {\bibnamefont {{J. Florencio}}}, \bibinfo {author}
  {\bibnamefont {{A. Saguia}}}, \ and\ \bibinfo {author} {\bibnamefont {{O. F.
  de Alcantara Bonfim}}},\ }\href@noop {} {\bibfield  {journal} {\bibinfo
  {journal} {Phys. Rev. E}\ }\textbf {\bibinfo {volume} {89}},\ \bibinfo
  {pages} {032143} (\bibinfo {year} {2014})}\BibitemShut {NoStop}%
\bibitem [{\citenamefont {{O. F. de Alcantara Bonfim}}\ \emph
  {et~al.}(2014)\citenamefont {{O. F. de Alcantara Bonfim}}, \citenamefont {{A.
  Saguia}}, \citenamefont {{B. Boechat}},\ and\ \citenamefont {{J.
  Florencio}}}]{ozi1}%
  \BibitemOpen
  \bibfield  {author} {\bibinfo {author} {\bibnamefont {{O. F. de Alcantara
  Bonfim}}}, \bibinfo {author} {\bibnamefont {{A. Saguia}}}, \bibinfo {author}
  {\bibnamefont {{B. Boechat}}}, \ and\ \bibinfo {author} {\bibnamefont {{J.
  Florencio}}},\ }\href@noop {} {\bibfield  {journal} {\bibinfo  {journal}
  {Phys. Rev. E}\ }\textbf {\bibinfo {volume} {90}},\ \bibinfo {pages} {032101}
  (\bibinfo {year} {2014})}\BibitemShut {NoStop}%
\bibitem [{\citenamefont {{O. F. de Alcantara Bonfim}}\ and\ \citenamefont {{J.
  Florencio}}(2006)}]{ozi2}%
  \BibitemOpen
  \bibfield  {author} {\bibinfo {author} {\bibnamefont {{O. F. de Alcantara
  Bonfim}}}\ and\ \bibinfo {author} {\bibnamefont {{J. Florencio}}},\
  }\href@noop {} {\bibfield  {journal} {\bibinfo  {journal} {Phys. Rev. B}\
  }\textbf {\bibinfo {volume} {74}},\ \bibinfo {pages} {134413} (\bibinfo
  {year} {2006})}\BibitemShut {NoStop}%
\bibitem [{\citenamefont {{G. T. Hohensee}}\ \emph {et~al.}(2014)\citenamefont
  {{G. T. Hohensee}}, \citenamefont {{R. B. Wilson}}, \citenamefont {{J. P.
  Feser}},\ and\ \citenamefont {{D. G. Cahill}}}]{hoh}%
  \BibitemOpen
  \bibfield  {author} {\bibinfo {author} {\bibnamefont {{G. T. Hohensee}}},
  \bibinfo {author} {\bibnamefont {{R. B. Wilson}}}, \bibinfo {author}
  {\bibnamefont {{J. P. Feser}}}, \ and\ \bibinfo {author} {\bibnamefont {{D.
  G. Cahill}}},\ }\href@noop {} {\bibfield  {journal} {\bibinfo  {journal}
  {Phys. Rev. B}\ }\textbf {\bibinfo {volume} {89}},\ \bibinfo {pages} {024422}
  (\bibinfo {year} {2014})}\BibitemShut {NoStop}%
\bibitem [{\citenamefont {{M. S. Naseri}}\ and\ \citenamefont {{S.
  Mahdavifar}}(2017)}]{nas}%
  \BibitemOpen
  \bibfield  {author} {\bibinfo {author} {\bibnamefont {{M. S. Naseri}}}\ and\
  \bibinfo {author} {\bibnamefont {{S. Mahdavifar}}},\ }\href@noop {}
  {\bibfield  {journal} {\bibinfo  {journal} {Physica A}\ }\textbf {\bibinfo
  {volume} {474}},\ \bibinfo {pages} {107} (\bibinfo {year}
  {2017})}\BibitemShut {NoStop}%
\bibitem [{\citenamefont {{A. Jabar}}\ \emph {et~al.}(2017)\citenamefont {{A.
  Jabar}}, \citenamefont {{N. Tahiri}}, \citenamefont {{K. Jetto}},\ and\
  \citenamefont {{L. Bahmad}}}]{jab}%
  \BibitemOpen
  \bibfield  {author} {\bibinfo {author} {\bibnamefont {{A. Jabar}}}, \bibinfo
  {author} {\bibnamefont {{N. Tahiri}}}, \bibinfo {author} {\bibnamefont {{K.
  Jetto}}}, \ and\ \bibinfo {author} {\bibnamefont {{L. Bahmad}}},\ }\href@noop
  {} {\bibfield  {journal} {\bibinfo  {journal} {Superlattices Microstruct.}\
  }\textbf {\bibinfo {volume} {104}},\ \bibinfo {pages} {46} (\bibinfo {year}
  {2017})}\BibitemShut {NoStop}%
\bibitem [{\citenamefont {{W. Chunle}}\ \emph {et~al.}(1988)\citenamefont {{W.
  Chunle}}, \citenamefont {{Q. Zikai}},\ and\ \citenamefont {{Z.
  Jinghbo}}}]{chu}%
  \BibitemOpen
  \bibfield  {author} {\bibinfo {author} {\bibnamefont {{W. Chunle}}}, \bibinfo
  {author} {\bibnamefont {{Q. Zikai}}}, \ and\ \bibinfo {author} {\bibnamefont
  {{Z. Jinghbo}}},\ }\href@noop {} {\bibfield  {journal} {\bibinfo  {journal}
  {Ferroelectrics}\ }\textbf {\bibinfo {volume} {77}},\ \bibinfo {pages} {21}
  (\bibinfo {year} {1988})}\BibitemShut {NoStop}%
\bibitem [{\citenamefont {{B. H. Teng}}\ and\ \citenamefont {{H. K.
  Sy}}(2006)}]{teng}%
  \BibitemOpen
  \bibfield  {author} {\bibinfo {author} {\bibnamefont {{B. H. Teng}}}\ and\
  \bibinfo {author} {\bibnamefont {{H. K. Sy}}},\ }\href@noop {} {\bibfield
  {journal} {\bibinfo  {journal} {Europhys. Lett.}\ }\textbf {\bibinfo {volume}
  {73}},\ \bibinfo {pages} {601} (\bibinfo {year} {2006})}\BibitemShut
  {NoStop}%
\bibitem [{\citenamefont {{P. Kowalewska}}\ and\ \citenamefont {{K.
  Szalowski}}(2020)}]{kowa}%
  \BibitemOpen
  \bibfield  {author} {\bibinfo {author} {\bibnamefont {{P. Kowalewska}}}\ and\
  \bibinfo {author} {\bibnamefont {{K. Szalowski}}},\ }\href@noop {} {\bibfield
   {journal} {\bibinfo  {journal} {J. Mag. Mag. Mat.}\ }\textbf {\bibinfo
  {volume} {496}},\ \bibinfo {pages} {165933} (\bibinfo {year}
  {2020})}\BibitemShut {NoStop}%
\bibitem [{\citenamefont {{J. Torrico}}\ and\ \citenamefont {{J. A.
  Plascak}}(2020)}]{jordana}%
  \BibitemOpen
  \bibfield  {author} {\bibinfo {author} {\bibnamefont {{J. Torrico}}}\ and\
  \bibinfo {author} {\bibnamefont {{J. A. Plascak}}},\ }\href@noop {}
  {\bibfield  {journal} {\bibinfo  {journal} {Phys. Rev. E}\ }\textbf {\bibinfo
  {volume} {102}},\ \bibinfo {pages} {062116} (\bibinfo {year}
  {2020})}\BibitemShut {NoStop}%
\bibitem [{\citenamefont {{K. Szalowski}}\ and\ \citenamefont {{P.
  Kowalewska}}(2020)}]{karol}%
  \BibitemOpen
  \bibfield  {author} {\bibinfo {author} {\bibnamefont {{K. Szalowski}}}\ and\
  \bibinfo {author} {\bibnamefont {{P. Kowalewska}}},\ }\href@noop {}
  {\bibfield  {journal} {\bibinfo  {journal} {Materials}\ }\textbf {\bibinfo
  {volume} {13}},\ \bibinfo {pages} {485} (\bibinfo {year} {2020})}\BibitemShut
  {NoStop}%
\bibitem [{\citenamefont {{S. Thomas}}\ \emph {et~al.}(2012)\citenamefont {{S.
  Thomas}}, \citenamefont {{S. Ramasesha}}, \citenamefont {{K. Hallberg}},\
  and\ \citenamefont {{D. Garcia}}}]{thomas}%
  \BibitemOpen
  \bibfield  {author} {\bibinfo {author} {\bibnamefont {{S. Thomas}}}, \bibinfo
  {author} {\bibnamefont {{S. Ramasesha}}}, \bibinfo {author} {\bibnamefont
  {{K. Hallberg}}}, \ and\ \bibinfo {author} {\bibnamefont {{D. Garcia}}},\
  }\href@noop {} {\bibfield  {journal} {\bibinfo  {journal} {Phys. Rev. B}\
  }\textbf {\bibinfo {volume} {86}},\ \bibinfo {pages} {180403} (\bibinfo
  {year} {2012})}\BibitemShut {NoStop}%
\bibitem [{\citenamefont {{A. Valentim}}\ \emph {et~al.}(2020)\citenamefont
  {{A. Valentim}}, \citenamefont {{G. A. Bocan}}, \citenamefont {{J. D. Fuhr}},
  \citenamefont {{D. J. Garcia}}, \citenamefont {{G. Giri}}, \citenamefont {{M.
  Kumar}},\ and\ \citenamefont {{S. Ramasesha}}}]{val}%
  \BibitemOpen
  \bibfield  {author} {\bibinfo {author} {\bibnamefont {{A. Valentim}}},
  \bibinfo {author} {\bibnamefont {{G. A. Bocan}}}, \bibinfo {author}
  {\bibnamefont {{J. D. Fuhr}}}, \bibinfo {author} {\bibnamefont {{D. J.
  Garcia}}}, \bibinfo {author} {\bibnamefont {{G. Giri}}}, \bibinfo {author}
  {\bibnamefont {{M. Kumar}}}, \ and\ \bibinfo {author} {\bibnamefont {{S.
  Ramasesha}}},\ }\href@noop {} {\bibfield  {journal} {\bibinfo  {journal}
  {Phys. Chem. Chem. Phys.}\ }\textbf {\bibinfo {volume} {22}},\ \bibinfo
  {pages} {5882} (\bibinfo {year} {2020})}\BibitemShut {NoStop}%
\bibitem [{\citenamefont {{C. Karrasch}}\ \emph {et~al.}(2015)\citenamefont
  {{C. Karrasch}}, \citenamefont {{D. M. Kennes}},\ and\ \citenamefont {{F.
  Heidrich-Meisner}}}]{kar}%
  \BibitemOpen
  \bibfield  {author} {\bibinfo {author} {\bibnamefont {{C. Karrasch}}},
  \bibinfo {author} {\bibnamefont {{D. M. Kennes}}}, \ and\ \bibinfo {author}
  {\bibnamefont {{F. Heidrich-Meisner}}},\ }\href@noop {} {\bibfield  {journal}
  {\bibinfo  {journal} {Phys. Rev. B}\ }\textbf {\bibinfo {volume} {91}},\
  \bibinfo {pages} {115130} (\bibinfo {year} {2015})}\BibitemShut {NoStop}%
\bibitem [{\citenamefont {{Z. Liu}}\ \emph {et~al.}(2017)\citenamefont {{Z.
  Liu}}, \citenamefont {{S. Jiang}}, \citenamefont {{X. Kong}},\ and\
  \citenamefont {{Y. Xu}}}]{liu}%
  \BibitemOpen
  \bibfield  {author} {\bibinfo {author} {\bibnamefont {{Z. Liu}}}, \bibinfo
  {author} {\bibnamefont {{S. Jiang}}}, \bibinfo {author} {\bibnamefont {{X.
  Kong}}}, \ and\ \bibinfo {author} {\bibnamefont {{Y. Xu}}},\ }\href@noop {}
  {\bibfield  {journal} {\bibinfo  {journal} {Physica A}\ }\textbf {\bibinfo
  {volume} {473}},\ \bibinfo {pages} {536} (\bibinfo {year}
  {2017})}\BibitemShut {NoStop}%
\bibitem [{\citenamefont {MacDonald}\ \emph {et~al.}(1988)\citenamefont
  {MacDonald}, \citenamefont {Girvin},\ and\ \citenamefont
  {Yoshioka}}]{macdonald}%
  \BibitemOpen
  \bibfield  {author} {\bibinfo {author} {\bibfnamefont {A.~H.}\ \bibnamefont
  {MacDonald}}, \bibinfo {author} {\bibfnamefont {S.~M.}\ \bibnamefont
  {Girvin}}, \ and\ \bibinfo {author} {\bibfnamefont {D.}~\bibnamefont
  {Yoshioka}},\ }\href@noop {} {\bibfield  {journal} {\bibinfo  {journal}
  {Phys. Rev. B}\ }\textbf {\bibinfo {volume} {37}},\ \bibinfo {pages} {9753}
  (\bibinfo {year} {1988})}\BibitemShut {NoStop}%
\bibitem [{\citenamefont {Motrunich}(2005)}]{motrunich}%
  \BibitemOpen
  \bibfield  {author} {\bibinfo {author} {\bibfnamefont {O.~I.}\ \bibnamefont
  {Motrunich}},\ }\href@noop {} {\bibfield  {journal} {\bibinfo  {journal}
  {Phys. Rev. B}\ }\textbf {\bibinfo {volume} {72}},\ \bibinfo {pages} {045105}
  (\bibinfo {year} {2005})}\BibitemShut {NoStop}%
\bibitem [{\citenamefont {Brehmer}\ \emph {et~al.}(1999)\citenamefont
  {Brehmer}, \citenamefont {Mikeska}, \citenamefont {M\"uller}, \citenamefont
  {Nagaosa},\ and\ \citenamefont {Uchida}}]{brehmer}%
  \BibitemOpen
  \bibfield  {author} {\bibinfo {author} {\bibfnamefont {S.}~\bibnamefont
  {Brehmer}}, \bibinfo {author} {\bibfnamefont {H.-J.}\ \bibnamefont
  {Mikeska}}, \bibinfo {author} {\bibfnamefont {M.}~\bibnamefont {M\"uller}},
  \bibinfo {author} {\bibfnamefont {N.}~\bibnamefont {Nagaosa}}, \ and\
  \bibinfo {author} {\bibfnamefont {S.}~\bibnamefont {Uchida}},\ }\href@noop {}
  {\bibfield  {journal} {\bibinfo  {journal} {Phys. Rev. B}\ }\textbf {\bibinfo
  {volume} {60}},\ \bibinfo {pages} {329} (\bibinfo {year} {1999})}\BibitemShut
  {NoStop}%
\bibitem [{\citenamefont {{M. Matsuda}}\ \emph {et~al.}(2000)\citenamefont {{M.
  Matsuda}}, \citenamefont {{K. Katsumata}}, \citenamefont {{R. S. Eccleston}},
  \citenamefont {{S. Brehmer}},\ and\ \citenamefont {{H.-J. Mikeska}}}]{mat}%
  \BibitemOpen
  \bibfield  {author} {\bibinfo {author} {\bibnamefont {{M. Matsuda}}},
  \bibinfo {author} {\bibnamefont {{K. Katsumata}}}, \bibinfo {author}
  {\bibnamefont {{R. S. Eccleston}}}, \bibinfo {author} {\bibnamefont {{S.
  Brehmer}}}, \ and\ \bibinfo {author} {\bibnamefont {{H.-J. Mikeska}}},\
  }\href@noop {} {\bibfield  {journal} {\bibinfo  {journal} {Phys. Rev. B}\
  }\textbf {\bibinfo {volume} {62}},\ \bibinfo {pages} {8903} (\bibinfo {year}
  {2000})}\BibitemShut {NoStop}%
\bibitem [{\citenamefont {{R. Coldea}}\ \emph {et~al.}(2001)\citenamefont {{R.
  Coldea}}, \citenamefont {{S. M. Hayden}}, \citenamefont {{G. Aeppli}},
  \citenamefont {{T. G. Perring}}, \citenamefont {{C. D. Frost}}, \citenamefont
  {{T. E. Mason}}, \citenamefont {{S.-W. Cheong}},\ and\ \citenamefont {{Z.
  Fisk}}}]{col}%
  \BibitemOpen
  \bibfield  {author} {\bibinfo {author} {\bibnamefont {{R. Coldea}}}, \bibinfo
  {author} {\bibnamefont {{S. M. Hayden}}}, \bibinfo {author} {\bibnamefont
  {{G. Aeppli}}}, \bibinfo {author} {\bibnamefont {{T. G. Perring}}}, \bibinfo
  {author} {\bibnamefont {{C. D. Frost}}}, \bibinfo {author} {\bibnamefont {{T.
  E. Mason}}}, \bibinfo {author} {\bibnamefont {{S.-W. Cheong}}}, \ and\
  \bibinfo {author} {\bibnamefont {{Z. Fisk}}},\ }\href@noop {} {\bibfield
  {journal} {\bibinfo  {journal} {Phys. Rev. Lett.}\ }\textbf {\bibinfo
  {volume} {86}},\ \bibinfo {pages} {5377} (\bibinfo {year}
  {2001})}\BibitemShut {NoStop}%
\bibitem [{\citenamefont {{C. B. Larsen}}\ \emph {et~al.}(2019)\citenamefont
  {{C. B. Larsen}}, \citenamefont {{A. T. Romer}}, \citenamefont {{S. Janas}},
  \citenamefont {{F. Treue}}, \citenamefont {{B. Monsted}}, \citenamefont {{N.
  E. Shaik}}, \citenamefont {{H. M. Ronnow}},\ and\ \citenamefont {{K.
  Lefmann}}}]{lar}%
  \BibitemOpen
  \bibfield  {author} {\bibinfo {author} {\bibnamefont {{C. B. Larsen}}},
  \bibinfo {author} {\bibnamefont {{A. T. Romer}}}, \bibinfo {author}
  {\bibnamefont {{S. Janas}}}, \bibinfo {author} {\bibnamefont {{F. Treue}}},
  \bibinfo {author} {\bibnamefont {{B. Monsted}}}, \bibinfo {author}
  {\bibnamefont {{N. E. Shaik}}}, \bibinfo {author} {\bibnamefont {{H. M.
  Ronnow}}}, \ and\ \bibinfo {author} {\bibnamefont {{K. Lefmann}}},\
  }\href@noop {} {\bibfield  {journal} {\bibinfo  {journal} {Phys. Rev. B}\
  }\textbf {\bibinfo {volume} {99}},\ \bibinfo {pages} {054432} (\bibinfo
  {year} {2019})}\BibitemShut {NoStop}%
\bibitem [{\citenamefont {Cookmeyer}\ \emph {et~al.}(2021)\citenamefont
  {Cookmeyer}, \citenamefont {Motruk},\ and\ \citenamefont
  {Moore}}]{Cookmeyer}%
  \BibitemOpen
  \bibfield  {author} {\bibinfo {author} {\bibfnamefont {T.}~\bibnamefont
  {Cookmeyer}}, \bibinfo {author} {\bibfnamefont {J.}~\bibnamefont {Motruk}}, \
  and\ \bibinfo {author} {\bibfnamefont {J.~E.}\ \bibnamefont {Moore}},\
  }\href@noop {} {\bibfield  {journal} {\bibinfo  {journal} {Phys. Rev. Lett.}\
  }\textbf {\bibinfo {volume} {127}},\ \bibinfo {pages} {087201} (\bibinfo
  {year} {2021})}\BibitemShut {NoStop}%
\bibitem [{\citenamefont {Chaboussant}\ \emph {et~al.}(1997)\citenamefont
  {Chaboussant}, \citenamefont {Crowell}, \citenamefont {L\'evy}, \citenamefont
  {Piovesana}, \citenamefont {Madouri},\ and\ \citenamefont
  {Mailly}}]{chaboussant}%
  \BibitemOpen
  \bibfield  {author} {\bibinfo {author} {\bibfnamefont {G.}~\bibnamefont
  {Chaboussant}}, \bibinfo {author} {\bibfnamefont {P.~A.}\ \bibnamefont
  {Crowell}}, \bibinfo {author} {\bibfnamefont {L.~P.}\ \bibnamefont {L\'evy}},
  \bibinfo {author} {\bibfnamefont {O.}~\bibnamefont {Piovesana}}, \bibinfo
  {author} {\bibfnamefont {A.}~\bibnamefont {Madouri}}, \ and\ \bibinfo
  {author} {\bibfnamefont {D.}~\bibnamefont {Mailly}},\ }\href@noop {}
  {\bibfield  {journal} {\bibinfo  {journal} {Phys. Rev. B}\ }\textbf {\bibinfo
  {volume} {55}},\ \bibinfo {pages} {3046} (\bibinfo {year}
  {1997})}\BibitemShut {NoStop}%
\bibitem [{\citenamefont {Watson}\ \emph {et~al.}(2001)\citenamefont {Watson},
  \citenamefont {Kotov}, \citenamefont {Meisel}, \citenamefont {Hall},
  \citenamefont {Granroth}, \citenamefont {Montfrooij}, \citenamefont {Nagler},
  \citenamefont {Jensen}, \citenamefont {Backov}, \citenamefont {Petruska},
  \citenamefont {Fanucci},\ and\ \citenamefont {Talham}}]{watson}%
  \BibitemOpen
  \bibfield  {author} {\bibinfo {author} {\bibfnamefont {B.~C.}\ \bibnamefont
  {Watson}}, \bibinfo {author} {\bibfnamefont {V.~N.}\ \bibnamefont {Kotov}},
  \bibinfo {author} {\bibfnamefont {M.~W.}\ \bibnamefont {Meisel}}, \bibinfo
  {author} {\bibfnamefont {D.~W.}\ \bibnamefont {Hall}}, \bibinfo {author}
  {\bibfnamefont {G.~E.}\ \bibnamefont {Granroth}}, \bibinfo {author}
  {\bibfnamefont {W.~T.}\ \bibnamefont {Montfrooij}}, \bibinfo {author}
  {\bibfnamefont {S.~E.}\ \bibnamefont {Nagler}}, \bibinfo {author}
  {\bibfnamefont {D.~A.}\ \bibnamefont {Jensen}}, \bibinfo {author}
  {\bibfnamefont {R.}~\bibnamefont {Backov}}, \bibinfo {author} {\bibfnamefont
  {M.~A.}\ \bibnamefont {Petruska}}, \bibinfo {author} {\bibfnamefont {G.~E.}\
  \bibnamefont {Fanucci}}, \ and\ \bibinfo {author} {\bibfnamefont {D.~R.}\
  \bibnamefont {Talham}},\ }\href@noop {} {\bibfield  {journal} {\bibinfo
  {journal} {Phys. Rev. Lett.}\ }\textbf {\bibinfo {volume} {86}},\ \bibinfo
  {pages} {5168} (\bibinfo {year} {2001})}\BibitemShut {NoStop}%
\bibitem [{\citenamefont {Hong}\ \emph {et~al.}(2010)\citenamefont {Hong},
  \citenamefont {Kim}, \citenamefont {Hotta}, \citenamefont {Takano},
  \citenamefont {Tremelling}, \citenamefont {Turnbull}, \citenamefont {Landee},
  \citenamefont {Kang}, \citenamefont {Christensen}, \citenamefont {Lefmann},
  \citenamefont {Schmidt}, \citenamefont {Uhrig},\ and\ \citenamefont
  {Broholm}}]{hong}%
  \BibitemOpen
  \bibfield  {author} {\bibinfo {author} {\bibfnamefont {T.}~\bibnamefont
  {Hong}}, \bibinfo {author} {\bibfnamefont {Y.~H.}\ \bibnamefont {Kim}},
  \bibinfo {author} {\bibfnamefont {C.}~\bibnamefont {Hotta}}, \bibinfo
  {author} {\bibfnamefont {Y.}~\bibnamefont {Takano}}, \bibinfo {author}
  {\bibfnamefont {G.}~\bibnamefont {Tremelling}}, \bibinfo {author}
  {\bibfnamefont {M.~M.}\ \bibnamefont {Turnbull}}, \bibinfo {author}
  {\bibfnamefont {C.~P.}\ \bibnamefont {Landee}}, \bibinfo {author}
  {\bibfnamefont {H.-J.}\ \bibnamefont {Kang}}, \bibinfo {author}
  {\bibfnamefont {N.~B.}\ \bibnamefont {Christensen}}, \bibinfo {author}
  {\bibfnamefont {K.}~\bibnamefont {Lefmann}}, \bibinfo {author} {\bibfnamefont
  {K.~P.}\ \bibnamefont {Schmidt}}, \bibinfo {author} {\bibfnamefont {G.~S.}\
  \bibnamefont {Uhrig}}, \ and\ \bibinfo {author} {\bibfnamefont
  {C.}~\bibnamefont {Broholm}},\ }\href@noop {} {\bibfield  {journal} {\bibinfo
   {journal} {Phys. Rev. Lett.}\ }\textbf {\bibinfo {volume} {105}},\ \bibinfo
  {pages} {137207} (\bibinfo {year} {2010})}\BibitemShut {NoStop}%
\bibitem [{\citenamefont {Schmidiger}\ \emph {et~al.}(2013)\citenamefont
  {Schmidiger}, \citenamefont {Bouillot}, \citenamefont {Guidi}, \citenamefont
  {Bewley}, \citenamefont {Kollath}, \citenamefont {Giamarchi},\ and\
  \citenamefont {Zheludev}}]{schmidiger}%
  \BibitemOpen
  \bibfield  {author} {\bibinfo {author} {\bibfnamefont {D.}~\bibnamefont
  {Schmidiger}}, \bibinfo {author} {\bibfnamefont {P.}~\bibnamefont
  {Bouillot}}, \bibinfo {author} {\bibfnamefont {T.}~\bibnamefont {Guidi}},
  \bibinfo {author} {\bibfnamefont {R.}~\bibnamefont {Bewley}}, \bibinfo
  {author} {\bibfnamefont {C.}~\bibnamefont {Kollath}}, \bibinfo {author}
  {\bibfnamefont {T.}~\bibnamefont {Giamarchi}}, \ and\ \bibinfo {author}
  {\bibfnamefont {A.}~\bibnamefont {Zheludev}},\ }\href@noop {} {\bibfield
  {journal} {\bibinfo  {journal} {Phys. Rev. Lett.}\ }\textbf {\bibinfo
  {volume} {111}},\ \bibinfo {pages} {107202} (\bibinfo {year}
  {2013})}\BibitemShut {NoStop}%
\bibitem [{\citenamefont {Blosser}\ \emph {et~al.}(2018)\citenamefont
  {Blosser}, \citenamefont {Bhartiya}, \citenamefont {Voneshen},\ and\
  \citenamefont {Zheludev}}]{blosser}%
  \BibitemOpen
  \bibfield  {author} {\bibinfo {author} {\bibfnamefont {D.}~\bibnamefont
  {Blosser}}, \bibinfo {author} {\bibfnamefont {V.~K.}\ \bibnamefont
  {Bhartiya}}, \bibinfo {author} {\bibfnamefont {D.~J.}\ \bibnamefont
  {Voneshen}}, \ and\ \bibinfo {author} {\bibfnamefont {A.}~\bibnamefont
  {Zheludev}},\ }\href@noop {} {\bibfield  {journal} {\bibinfo  {journal}
  {Phys. Rev. Lett.}\ }\textbf {\bibinfo {volume} {121}},\ \bibinfo {pages}
  {247201} (\bibinfo {year} {2018})}\BibitemShut {NoStop}%
\bibitem [{\citenamefont {{D. C. Cabra}}\ \emph {et~al.}(1997)\citenamefont
  {{D. C. Cabra}}, \citenamefont {{A. Honecker}},\ and\ \citenamefont {{P.
  Pujol}}}]{cabraetal0}%
  \BibitemOpen
  \bibfield  {author} {\bibinfo {author} {\bibnamefont {{D. C. Cabra}}},
  \bibinfo {author} {\bibnamefont {{A. Honecker}}}, \ and\ \bibinfo {author}
  {\bibnamefont {{P. Pujol}}},\ }\href@noop {} {\bibfield  {journal} {\bibinfo
  {journal} {Phys. Rev. Lett.}\ }\textbf {\bibinfo {volume} {79}},\ \bibinfo
  {pages} {5126} (\bibinfo {year} {1997})}\BibitemShut {NoStop}%
\bibitem [{\citenamefont {Honecker}\ \emph {et~al.}(2000)\citenamefont
  {Honecker}, \citenamefont {Mila},\ and\ \citenamefont
  {Troyer}}]{honecker2000}%
  \BibitemOpen
  \bibfield  {author} {\bibinfo {author} {\bibfnamefont {A.}~\bibnamefont
  {Honecker}}, \bibinfo {author} {\bibfnamefont {F.}~\bibnamefont {Mila}}, \
  and\ \bibinfo {author} {\bibfnamefont {M.}~\bibnamefont {Troyer}},\
  }\href@noop {} {\bibfield  {journal} {\bibinfo  {journal} {Eur. Phys. J. B.}\
  }\textbf {\bibinfo {volume} {15}},\ \bibinfo {pages} {227} (\bibinfo {year}
  {2000})}\BibitemShut {NoStop}%
\bibitem [{\citenamefont {Fouet}\ \emph {et~al.}(2006)\citenamefont {Fouet},
  \citenamefont {Mila}, \citenamefont {Clarke}, \citenamefont {Youk},
  \citenamefont {Tchernyshyov}, \citenamefont {Fendley},\ and\ \citenamefont
  {Noack}}]{fouet}%
  \BibitemOpen
  \bibfield  {author} {\bibinfo {author} {\bibfnamefont {J.-B.}\ \bibnamefont
  {Fouet}}, \bibinfo {author} {\bibfnamefont {F.}~\bibnamefont {Mila}},
  \bibinfo {author} {\bibfnamefont {D.}~\bibnamefont {Clarke}}, \bibinfo
  {author} {\bibfnamefont {H.}~\bibnamefont {Youk}}, \bibinfo {author}
  {\bibfnamefont {O.}~\bibnamefont {Tchernyshyov}}, \bibinfo {author}
  {\bibfnamefont {P.}~\bibnamefont {Fendley}}, \ and\ \bibinfo {author}
  {\bibfnamefont {R.~M.}\ \bibnamefont {Noack}},\ }\href@noop {} {\bibfield
  {journal} {\bibinfo  {journal} {Phys. Rev. B}\ }\textbf {\bibinfo {volume}
  {73}},\ \bibinfo {pages} {214405} (\bibinfo {year} {2006})}\BibitemShut
  {NoStop}%
\bibitem [{\citenamefont {Starykh}\ and\ \citenamefont
  {Balents}(2004)}]{Starykh2004}%
  \BibitemOpen
  \bibfield  {author} {\bibinfo {author} {\bibfnamefont {O.~A.}\ \bibnamefont
  {Starykh}}\ and\ \bibinfo {author} {\bibfnamefont {L.}~\bibnamefont
  {Balents}},\ }\href@noop {} {\bibfield  {journal} {\bibinfo  {journal} {Phys.
  Rev. Lett.}\ }\textbf {\bibinfo {volume} {93}},\ \bibinfo {pages} {127202}
  (\bibinfo {year} {2004})}\BibitemShut {NoStop}%
\bibitem [{\citenamefont {Hikihara}\ and\ \citenamefont
  {Starykh}(2010)}]{Hikihara}%
  \BibitemOpen
  \bibfield  {author} {\bibinfo {author} {\bibfnamefont {T.}~\bibnamefont
  {Hikihara}}\ and\ \bibinfo {author} {\bibfnamefont {O.~A.}\ \bibnamefont
  {Starykh}},\ }\href@noop {} {\bibfield  {journal} {\bibinfo  {journal} {Phys.
  Rev. B}\ }\textbf {\bibinfo {volume} {81}},\ \bibinfo {pages} {064432}
  (\bibinfo {year} {2010})}\BibitemShut {NoStop}%
\bibitem [{\citenamefont {Barcza}\ \emph {et~al.}(2012)\citenamefont {Barcza},
  \citenamefont {Legeza}, \citenamefont {Noack},\ and\ \citenamefont
  {S\'olyom}}]{Barcza}%
  \BibitemOpen
  \bibfield  {author} {\bibinfo {author} {\bibfnamefont {G.}~\bibnamefont
  {Barcza}}, \bibinfo {author} {\bibfnamefont {O.}~\bibnamefont {Legeza}},
  \bibinfo {author} {\bibfnamefont {R.~M.}\ \bibnamefont {Noack}}, \ and\
  \bibinfo {author} {\bibfnamefont {J.}~\bibnamefont {S\'olyom}},\ }\href@noop
  {} {\bibfield  {journal} {\bibinfo  {journal} {Phys. Rev. B}\ }\textbf
  {\bibinfo {volume} {86}},\ \bibinfo {pages} {075133} (\bibinfo {year}
  {2012})}\BibitemShut {NoStop}%
\bibitem [{\citenamefont {Sakai}\ and\ \citenamefont {Hasegawa}(1999)}]{sakai}%
  \BibitemOpen
  \bibfield  {author} {\bibinfo {author} {\bibfnamefont {T.}~\bibnamefont
  {Sakai}}\ and\ \bibinfo {author} {\bibfnamefont {Y.}~\bibnamefont
  {Hasegawa}},\ }\href@noop {} {\bibfield  {journal} {\bibinfo  {journal}
  {Phys. Rev. B}\ }\textbf {\bibinfo {volume} {60}},\ \bibinfo {pages} {48}
  (\bibinfo {year} {1999})}\BibitemShut {NoStop}%
\bibitem [{\citenamefont {L\"auchli}\ \emph {et~al.}(2003)\citenamefont
  {L\"auchli}, \citenamefont {Schmid},\ and\ \citenamefont {Troyer}}]{lauchli}%
  \BibitemOpen
  \bibfield  {author} {\bibinfo {author} {\bibfnamefont {A.}~\bibnamefont
  {L\"auchli}}, \bibinfo {author} {\bibfnamefont {G.}~\bibnamefont {Schmid}}, \
  and\ \bibinfo {author} {\bibfnamefont {M.}~\bibnamefont {Troyer}},\
  }\href@noop {} {\bibfield  {journal} {\bibinfo  {journal} {Phys. Rev. B}\
  }\textbf {\bibinfo {volume} {67}},\ \bibinfo {pages} {100409} (\bibinfo
  {year} {2003})}\BibitemShut {NoStop}%
\bibitem [{\citenamefont {Calzado}\ \emph {et~al.}(2003)\citenamefont
  {Calzado}, \citenamefont {de~Graaf}, \citenamefont {Bordas}, \citenamefont
  {Caballol},\ and\ \citenamefont {Malrieu}}]{calzado}%
  \BibitemOpen
  \bibfield  {author} {\bibinfo {author} {\bibfnamefont {C.~J.}\ \bibnamefont
  {Calzado}}, \bibinfo {author} {\bibfnamefont {C.}~\bibnamefont {de~Graaf}},
  \bibinfo {author} {\bibfnamefont {E.}~\bibnamefont {Bordas}}, \bibinfo
  {author} {\bibfnamefont {R.}~\bibnamefont {Caballol}}, \ and\ \bibinfo
  {author} {\bibfnamefont {J.-P.}\ \bibnamefont {Malrieu}},\ }\href@noop {}
  {\bibfield  {journal} {\bibinfo  {journal} {Phys. Rev. B}\ }\textbf {\bibinfo
  {volume} {67}},\ \bibinfo {pages} {132409} (\bibinfo {year}
  {2003})}\BibitemShut {NoStop}%
\bibitem [{\citenamefont {Gritsev}\ \emph {et~al.}(2004)\citenamefont
  {Gritsev}, \citenamefont {Normand},\ and\ \citenamefont
  {Baeriswyl}}]{gritsev}%
  \BibitemOpen
  \bibfield  {author} {\bibinfo {author} {\bibfnamefont {V.}~\bibnamefont
  {Gritsev}}, \bibinfo {author} {\bibfnamefont {B.}~\bibnamefont {Normand}}, \
  and\ \bibinfo {author} {\bibfnamefont {D.}~\bibnamefont {Baeriswyl}},\
  }\href@noop {} {\bibfield  {journal} {\bibinfo  {journal} {Phys. Rev. B}\
  }\textbf {\bibinfo {volume} {69}},\ \bibinfo {pages} {094431} (\bibinfo
  {year} {2004})}\BibitemShut {NoStop}%
\bibitem [{\citenamefont {Liu}\ \emph {et~al.}(2008)\citenamefont {Liu},
  \citenamefont {Wang},\ and\ \citenamefont {Tian}}]{liu2008}%
  \BibitemOpen
  \bibfield  {author} {\bibinfo {author} {\bibfnamefont {G.-H.}\ \bibnamefont
  {Liu}}, \bibinfo {author} {\bibfnamefont {H.-L.}\ \bibnamefont {Wang}}, \
  and\ \bibinfo {author} {\bibfnamefont {G.-S.}\ \bibnamefont {Tian}},\
  }\href@noop {} {\bibfield  {journal} {\bibinfo  {journal} {Phys. Rev. B}\
  }\textbf {\bibinfo {volume} {77}},\ \bibinfo {pages} {214418} (\bibinfo
  {year} {2008})}\BibitemShut {NoStop}%
\bibitem [{\citenamefont {{S. Capponi}}\ \emph {et~al.}(2013)\citenamefont {{S.
  Capponi}}, \citenamefont {{P. Lecheminant}},\ and\ \citenamefont {{M.
  Moliner}}}]{caponi-nleg}%
  \BibitemOpen
  \bibfield  {author} {\bibinfo {author} {\bibnamefont {{S. Capponi}}},
  \bibinfo {author} {\bibnamefont {{P. Lecheminant}}}, \ and\ \bibinfo {author}
  {\bibnamefont {{M. Moliner}}},\ }\href@noop {} {\bibfield  {journal}
  {\bibinfo  {journal} {Phys. Rev. B}\ }\textbf {\bibinfo {volume} {88}},\
  \bibinfo {pages} {075132} (\bibinfo {year} {2013})}\BibitemShut {NoStop}%
\bibitem [{\citenamefont {Kohmoto}\ \emph {et~al.}(1981)\citenamefont
  {Kohmoto}, \citenamefont {den Nijs},\ and\ \citenamefont
  {Kadanoff}}]{kohmoto}%
  \BibitemOpen
  \bibfield  {author} {\bibinfo {author} {\bibfnamefont {M.}~\bibnamefont
  {Kohmoto}}, \bibinfo {author} {\bibfnamefont {M.}~\bibnamefont {den Nijs}}, \
  and\ \bibinfo {author} {\bibfnamefont {L.~P.}\ \bibnamefont {Kadanoff}},\
  }\href@noop {} {\bibfield  {journal} {\bibinfo  {journal} {Phys. Rev. B}\
  }\textbf {\bibinfo {volume} {24}},\ \bibinfo {pages} {5229} (\bibinfo {year}
  {1981})}\BibitemShut {NoStop}%
\bibitem [{\citenamefont {{F. C. Alcaraz}}\ \emph {et~al.}(1987)\citenamefont
  {{F. C. Alcaraz}}, \citenamefont {{M. N. Barber}},\ and\ \citenamefont {{M.
  T. Batchelor}}}]{quantumAsh-Teller}%
  \BibitemOpen
  \bibfield  {author} {\bibinfo {author} {\bibnamefont {{F. C. Alcaraz}}},
  \bibinfo {author} {\bibnamefont {{M. N. Barber}}}, \ and\ \bibinfo {author}
  {\bibnamefont {{M. T. Batchelor}}},\ }\href@noop {} {\bibfield  {journal}
  {\bibinfo  {journal} {Phys. Rev. Lett.}\ }\textbf {\bibinfo {volume} {58}},\
  \bibinfo {pages} {771} (\bibinfo {year} {1987})}\BibitemShut {NoStop}%
\bibitem [{\citenamefont {{J. Eisert}}\ \emph {et~al.}(2010)\citenamefont {{J.
  Eisert}}, \citenamefont {{M. Cramer}},\ and\ \citenamefont {{M. B.
  Plenio}}}]{RMP82-277}%
  \BibitemOpen
  \bibfield  {author} {\bibinfo {author} {\bibnamefont {{J. Eisert}}}, \bibinfo
  {author} {\bibnamefont {{M. Cramer}}}, \ and\ \bibinfo {author} {\bibnamefont
  {{M. B. Plenio}}},\ }\href@noop {} {\bibfield  {journal} {\bibinfo  {journal}
  {Rev. Mod. Phys.}\ }\textbf {\bibinfo {volume} {82}},\ \bibinfo {pages} {277}
  (\bibinfo {year} {2010})}\BibitemShut {NoStop}%
\bibitem [{\citenamefont {{C. Holzhey}}\ \emph {et~al.}(1994)\citenamefont {{C.
  Holzhey}}, \citenamefont {{F. Larsen}},\ and\ \citenamefont {{F.
  Wilczek}}}]{cold}%
  \BibitemOpen
  \bibfield  {author} {\bibinfo {author} {\bibnamefont {{C. Holzhey}}},
  \bibinfo {author} {\bibnamefont {{F. Larsen}}}, \ and\ \bibinfo {author}
  {\bibnamefont {{F. Wilczek}}},\ }\href@noop {} {\bibfield  {journal}
  {\bibinfo  {journal} {Nucl. Phys.}\ }\textbf {\bibinfo {volume} {B424}},\
  \bibinfo {pages} {443} (\bibinfo {year} {1994})}\BibitemShut {NoStop}%
\bibitem [{\citenamefont {{P. Calabrese}}\ and\ \citenamefont {{J.
  Cardy}}(2004)}]{cardyentan}%
  \BibitemOpen
  \bibfield  {author} {\bibinfo {author} {\bibnamefont {{P. Calabrese}}}\ and\
  \bibinfo {author} {\bibnamefont {{J. Cardy}}},\ }\href@noop {} {\bibfield
  {journal} {\bibinfo  {journal} {J. Stat. Mech.}\ ,\ \bibinfo {pages}
  {P06002}} (\bibinfo {year} {2004})}\BibitemShut {NoStop}%
\bibitem [{\citenamefont {{P. Calabrese}}\ and\ \citenamefont {{J.
  Cardy}}(2009)}]{entroreviewcalabrese}%
  \BibitemOpen
  \bibfield  {author} {\bibinfo {author} {\bibnamefont {{P. Calabrese}}}\ and\
  \bibinfo {author} {\bibnamefont {{J. Cardy}}},\ }\href@noop {} {\bibfield
  {journal} {\bibinfo  {journal} {J. Phys. A: Math. Theor.}\ }\textbf {\bibinfo
  {volume} {42}},\ \bibinfo {pages} {504005} (\bibinfo {year}
  {2009})}\BibitemShut {NoStop}%
\bibitem [{\citenamefont {{I. Affleck}}\ and\ \citenamefont {{A. W. W.
  Ludwig}}(1991)}]{affleckboundary}%
  \BibitemOpen
  \bibfield  {author} {\bibinfo {author} {\bibnamefont {{I. Affleck}}}\ and\
  \bibinfo {author} {\bibnamefont {{A. W. W. Ludwig}}},\ }\href@noop {}
  {\bibfield  {journal} {\bibinfo  {journal} {Phys. Rev. Lett.}\ }\textbf
  {\bibinfo {volume} {67}},\ \bibinfo {pages} {161} (\bibinfo {year}
  {1991})}\BibitemShut {NoStop}%
\bibitem [{\citenamefont {{P. Calabrese}}\ \emph {et~al.}(2010)\citenamefont
  {{P. Calabrese}}, \citenamefont {{M. Campostrini}}, \citenamefont {{F.
  Essler}},\ and\ \citenamefont {{B. Nienhuis}}}]{entropyosc}%
  \BibitemOpen
  \bibfield  {author} {\bibinfo {author} {\bibnamefont {{P. Calabrese}}},
  \bibinfo {author} {\bibnamefont {{M. Campostrini}}}, \bibinfo {author}
  {\bibnamefont {{F. Essler}}}, \ and\ \bibinfo {author} {\bibnamefont {{B.
  Nienhuis}}},\ }\href@noop {} {\bibfield  {journal} {\bibinfo  {journal}
  {Phys. Rev. Lett.}\ }\textbf {\bibinfo {volume} {104}},\ \bibinfo {pages}
  {095701} (\bibinfo {year} {2010})}\BibitemShut {NoStop}%
\bibitem [{\citenamefont {{P. Calabrese}}\ and\ \citenamefont {{F. H. L.
  Essler}}(2010)}]{xxPBCh}%
  \BibitemOpen
  \bibfield  {author} {\bibinfo {author} {\bibnamefont {{P. Calabrese}}}\ and\
  \bibinfo {author} {\bibnamefont {{F. H. L. Essler}}},\ }\href@noop {}
  {\bibfield  {journal} {\bibinfo  {journal} {J. Stat. Mech.}\ ,\ \bibinfo
  {pages} {P08029}} (\bibinfo {year} {2010})}\BibitemShut {NoStop}%
\bibitem [{\citenamefont {{M. Fagotti}}\ and\ \citenamefont {{P.
  Calabrese}}(2011)}]{calabreseOBC}%
  \BibitemOpen
  \bibfield  {author} {\bibinfo {author} {\bibnamefont {{M. Fagotti}}}\ and\
  \bibinfo {author} {\bibnamefont {{P. Calabrese}}},\ }\href@noop {} {\bibfield
   {journal} {\bibinfo  {journal} {J. Stat. Mech.}\ ,\ \bibinfo {pages}
  {P01017}} (\bibinfo {year} {2011})}\BibitemShut {NoStop}%
\bibitem [{\citenamefont {{J. C. Xavier}}\ and\ \citenamefont {{F. C.
  Alcaraz}}(2011{\natexlab{a}})}]{xavieralcarazosc}%
  \BibitemOpen
  \bibfield  {author} {\bibinfo {author} {\bibnamefont {{J. C. Xavier}}}\ and\
  \bibinfo {author} {\bibnamefont {{F. C. Alcaraz}}},\ }\href@noop {}
  {\bibfield  {journal} {\bibinfo  {journal} {Phys. Rev. B}\ }\textbf {\bibinfo
  {volume} {83}},\ \bibinfo {pages} {214425} (\bibinfo {year}
  {2011}{\natexlab{a}})}\BibitemShut {NoStop}%
\bibitem [{\citenamefont {{J. C. Xavier}}\ and\ \citenamefont {{F. C.
  Alcaraz}}(2012)}]{XavierAlca2012}%
  \BibitemOpen
  \bibfield  {author} {\bibinfo {author} {\bibnamefont {{J. C. Xavier}}}\ and\
  \bibinfo {author} {\bibnamefont {{F. C. Alcaraz}}},\ }\href@noop {}
  {\bibfield  {journal} {\bibinfo  {journal} {Phys. Rev. B}\ }\textbf {\bibinfo
  {volume} {85}},\ \bibinfo {pages} {024418} (\bibinfo {year}
  {2012})}\BibitemShut {NoStop}%
\bibitem [{\citenamefont {{J. Cardy}}\ and\ \citenamefont {{P.
  Calabrese}}(2010)}]{cardyosc}%
  \BibitemOpen
  \bibfield  {author} {\bibinfo {author} {\bibnamefont {{J. Cardy}}}\ and\
  \bibinfo {author} {\bibnamefont {{P. Calabrese}}},\ }\href@noop {} {\bibfield
   {journal} {\bibinfo  {journal} {J. Stat. Mech.}\ ,\ \bibinfo {pages}
  {P04023}} (\bibinfo {year} {2010})}\BibitemShut {NoStop}%
\bibitem [{\citenamefont {{J. C. Xavier}}\ and\ \citenamefont {{F. C.
  Alcaraz}}(2011{\natexlab{b}})}]{xavieralcarazQCP}%
  \BibitemOpen
  \bibfield  {author} {\bibinfo {author} {\bibnamefont {{J. C. Xavier}}}\ and\
  \bibinfo {author} {\bibnamefont {{F. C. Alcaraz}}},\ }\href@noop {}
  {\bibfield  {journal} {\bibinfo  {journal} {Phys. Rev. B}\ }\textbf {\bibinfo
  {volume} {84}},\ \bibinfo {pages} {094410} (\bibinfo {year}
  {2011}{\natexlab{b}})}\BibitemShut {NoStop}%
\bibitem [{\citenamefont {{J. C. Xavier}}\ and\ \citenamefont {{F. B.
  Ramos}}(2014)}]{arealawflavia}%
  \BibitemOpen
  \bibfield  {author} {\bibinfo {author} {\bibnamefont {{J. C. Xavier}}}\ and\
  \bibinfo {author} {\bibnamefont {{F. B. Ramos}}},\ }\href@noop {} {\bibfield
  {journal} {\bibinfo  {journal} {J. Stat. Mech}\ ,\ \bibinfo {pages} {P10034}}
  (\bibinfo {year} {2014})}\BibitemShut {NoStop}%
\bibitem [{\citenamefont {Gogolin}\ \emph {et~al.}(1998)\citenamefont
  {Gogolin}, \citenamefont {Nersesyan},\ and\ \citenamefont
  {Tsvelik}}]{Gogolin}%
  \BibitemOpen
  \bibfield  {author} {\bibinfo {author} {\bibfnamefont {A.~O.}\ \bibnamefont
  {Gogolin}}, \bibinfo {author} {\bibfnamefont {A.~A.}\ \bibnamefont
  {Nersesyan}}, \ and\ \bibinfo {author} {\bibfnamefont {A.~M.}\ \bibnamefont
  {Tsvelik}},\ }\href@noop {} {\emph {\bibinfo {title} {Bosonization and
  Strongly Correlated System}}}\ (\bibinfo  {publisher} {Cambridge University
  Press},\ \bibinfo {year} {1998})\BibitemShut {NoStop}%
\bibitem [{\citenamefont {{P. Di Francesco}}\ \emph {et~al.}(1999)\citenamefont
  {{P. Di Francesco}}, \citenamefont {{P. Mathieu}},\ and\ \citenamefont {{D.
  Senechal}}}]{critical3}%
  \BibitemOpen
  \bibfield  {author} {\bibinfo {author} {\bibnamefont {{P. Di Francesco}}},
  \bibinfo {author} {\bibnamefont {{P. Mathieu}}}, \ and\ \bibinfo {author}
  {\bibnamefont {{D. Senechal}}},\ }\href@noop {} {\emph {\bibinfo {title}
  {Conformal {F}ield {T}heory}}}\ (\bibinfo  {publisher} {Springer},\ \bibinfo
  {address} {New York},\ \bibinfo {year} {1999})\BibitemShut {NoStop}%
\end{thebibliography}%

\end{document}